\documentclass[journal=jctcce,manuscript=article]{achemso}

\usepackage[version=3]{mhchem} 
\usepackage{graphicx}
\usepackage[caption=false]{subfig} 
\usepackage{amssymb}
\usepackage{amsmath}
\usepackage{bm}
\usepackage{color}

\usepackage{natbib} 
\usepackage[resetlabels]{multibib}
\newcites{sm}{sm}
 \usepackage{float}
 
\usepackage{algorithm,algpseudocode}
\algnewcommand{\To}{\textbf{To }}
\algnewcommand\Input{\item[\textbf{Input:}]}%
\algnewcommand\Output{\item[\textbf{Output:}]}%

\title[An \textsf{achemso} demo]
  {Efficient Irreversible Monte Carlo Samplers}

\author{Fahim Faizi}
\affiliation[King's College London]
{Department of Mathematics, King's College London, Strand WC2R 2LS, London, U.K.}
\email{fahim.faizi@kcl.ac.uk}

\author{George Deligiannidis}
\affiliation[University of Oxford]
{Department of Statistics, University of Oxford, 24-29 St Giles', OX1 3LB, Oxford, U.K.}
\email{deligian@stats.ox.ac.uk}

\author{Edina Rosta}
\affiliation[King's college London]
{Department of Chemistry, King's College London, 7 Trinity street,  SE1 1DB, London, U.K.}
\email{edina.rosta@kcl.ac.uk}

\begin{document}
  
\begin{abstract}
  We present here two irreversible Markov chain Monte Carlo algorithms for general discrete state systems, one of the algorithms is based on the random-scan Gibbs sampler for discrete states and the other on its improved version, the Metropolized-Gibbs sampler. The algorithms we present incorporate the lifting framework with skewed detailed balance condition and construct irreversible Markov chains that satisfy the balance condition. We have applied our algorithms to 1D 4-state Potts model. The integrated autocorrelation times for magnetisation and energy density indicate a reduction of the dynamical scaling exponent from $z \approx 1$ to $z \approx  1/2$. In addition, we have generalized an irreversible Metropolis-Hastings algorithm with skewed detailed balance, initially introduced by Turitsyn et al. \cite{Turitsyn} for the mean field Ising model, to be now readily applicable to classical spin systems in general; application to 1D 4-state Potts model indicate a square root reduction of the mixing time at high temperatures.
\end{abstract}

\section*{Introduction}
Markov Chain Monte Carlo methods (MCMC) have been extensively utilised to the investigation of a broad range of statistical problems encompassing Physics \cite{MCMC physics1, MCMC physics2}, Biochemical sciences \cite{MCMC biochemistry, MCMC biochemistry2} and Economics and Finance \cite{MCMC finance}. The introduction of the widely used Metropolis algorithm \cite{Metropolis} in 1953 paved the path to a broad application of MCMC methods in sampling from probability distributions with very large dimensions, mostly with the ultimate intention to estimate expectation values of observables under such distributions. 

In order to ensure sampling from the desired distribution MCMC methods require the propagation of a Markov chain by a carefully constructed transition probability such that the invariant distribution of the Markov chain is precisely a desired target distribution. Conventional MCMC methods in statistical physics such as the Metropolis criteria and the Gibbs sampler \cite{Gibbs sampler} impose the strict detailed balance condition (DBC) on the transition matrix to ensure sampling from the desired distribution, in addition all MCMC methods must impose ergodicity to ensure convergence to the invariant.

In the DBC regime, where every elementary transition must balance with its corresponding inverse process, several improvements on the Metropolis Monte Carlo methods have been proposed to boost relaxation times. One such category is the generalized-ensemble algorithms \cite{Generalized ensemble}, common examples of which include the parallel tempering \cite{REM 1, REM 3, REM 4, REM 5}, simulated tempering \cite{STM} and multi-canonical methods \cite{MUCA 1, MUCA 2}, these algorithms have been very successful in simulation of complex bio-molecular systems with many degrees of freedom and a large number of local minimum energy states. Another class of algorithms with DBC are the cluster algorithms in classical spin systems such as the Swendsen-Wang \cite{Swendsen and Wang} and Wolff algorithm \cite{Wolff}, whereby the multi-spin update through a careful construction of a transition matrix drastically reduces the critical slowing down \cite{MCMC physics2} of spin systems.

In the DBC regime Peskun's theorem \cite{Peskun} dictates that the asymptotic variance on a given observable is reduced by the minimisation of the rejection rate in the Markov chain. Liu \cite{Liu1, Liu2} has successfully applied this idea to the random scan Gibbs sampler (GS) on discrete state spaces to construct the \textit{Metropolized-Gibbs} sampler (MGS) which yields smaller diagonal elements in the transition matrix \cite{LOU}. Pollet et al. have applied MGS to $q$ = 4 state Potts model \cite{Metropolized-Gibbs} where compared to the random scan Gibbs sampler a reduction in the asymptotic variance on the energy of the system is achieved at the critical temperature.

The strict detailed balance condition is however not a necessary requirement to ensure the invariance of the target distribution, the more general balance condition (BC) is mathematically sufficient \cite{BC sufficiency1, BC sufficiency2, BC sufficiency3}. The violation of DBC to improve sampling efficiency of MCMC algorithms has been a hot topic of discussion in various scenarios \cite{hot topic 1, hot topic 2, hot topic 3, hot topic 4, hot topic 5, hot topic 6, hot topic 7} with several numerical and analytical studies demonstrating improved sampling efficiency of MCMC methods that violate DBC but satisfy BC to ensure invariance \cite{Diaconis, Chen, Barkema, Ren, Suwa-Todo, Suwa-Todo2, Turitsyn, Weigel, Sakai Hukushima 1D, Sakai Hukushima 2D, Sakai Hukushima eigenvalue, Sakai Hukushima simulated tempering, ECMC continuous spins, ECMC hard spheres, ECMC generalized, ECMC heisenberg, non reversible parallel tempering}. 

There are various methods of violating DBC. For a classical spin system with local spin updating the random updating scheme, whereby a spin is chosen at random, satisfies DBC, whereas the sequential updating scheme, whereby spins are updated in a sequential order (e.g., in one sweep), satisfies DBC only locally (i.e., only at each spin flip). The transition kernel of each sweep, however, breaks DBC but satisfies BC to ensure invariance \cite{BC sufficiency3, Ren}. 

Suwa and Todo have proposed a novel method based on geometric weight allocation which satisfies BC but violates DBC even locally \cite{Suwa-Todo, Suwa-Todo2}. The authors have applied their algorithm to $q$ = 4 and 8 state Potts model reporting a boost in the relaxation time in both cases compared to the Metropolis-Hastings algorithm - by a factor of 6.4 for 4-state Potts model. The Suwa-Todo algorithm has since been extended to generalized-ensemble algorithms such as simulated tempering \cite{ST Suwa-Todo} and replica permutation method \cite{REM ST1, REM ST2}.

Another class of irreversible methods that have been an eager topic of study incorporate the concept of lifting \cite{Diaconis, Chen, Barkema, Turitsyn, Weigel, Sakai Hukushima 1D, Sakai Hukushima 2D, Sakai Hukushima eigenvalue, Sakai Hukushima simulated tempering, ECMC continuous spins, ECMC hard spheres, ECMC generalized, ECMC heisenberg, non reversible parallel tempering}. In the lifting framework of Diaconis et al. \cite{Diaconis} the state space and the target distribution are extended by creating a duplicate replica of the system, each replica characterised by a lifting variable, and each state in the state space therefore acquiring two copies, one in each replica. An irreversible lifted Markov chain is thus propagated in this enlarged state space by a transition matrix that violates DBC but yet ensures invariance of the target distribution by satisfying BC. The lifting framework has been applied to mean-field Ising model \cite{Turitsyn, Weigel}, where the integrated autocorrelation time of magnetisation reportedly indicates a reduction in the dynamical scaling exponent at the critical temperature. 

To augment the state space the lifting mechanism has been incorporated in event-chain Monte Carlo algorithms (ECMC) \cite{ ECMC hard spheres}, initially constructed for hard disk and hard sphere systems and later adapted for more general particle systems with continuous degrees of freedom \cite{ECMC generalized}. Further applications of ECMC with the lifting mechanism to continuous spin systems such as the three dimensional Heisenberg model has led to $ z \simeq 1$ dynamic scaling \cite{ECMC heisenberg}, while a speed up by two orders is reported with respect to local Metropolis MC in the autocorrelation time for magnetic susceptibility for the XY model \cite{ECMC continuous spins}.

The research presented in this paper concerns the framework of lifting with the skewed detailed balance condition (SDBC), originally proposed by Turitsyn et al. \cite{Turitsyn} and extensively studied by Sakai and Hukushima \cite{Sakai Hukushima 1D, Sakai Hukushima 2D, Sakai Hukushima eigenvalue, Sakai Hukushima simulated tempering}. Our work here is particularly motivated by the analytical and numerical studies of irreversible Glauber dynamics with SDBC for the cases of one  and two dimensional Ising model \cite{Sakai Hukushima 1D, Sakai Hukushima 2D}. In this paper we present two main generalizations of the works of Turitsyn et al.\cite{Turitsyn} and  Sakai and Hukushima \cite{Sakai Hukushima 1D, Sakai Hukushima 2D, Sakai Hukushima eigenvalue}: \textbf{1:} We have generalized an irreversible Metropolis-Hastings algorithm (IMH) with SDBC for the Ising model \cite{Turitsyn, Sakai Hukushima eigenvalue} to be now readily applicable to classical spin systems in general. \textbf{2:} Using the same lifting technique of Turitsyn et al. \cite{Turitsyn} we have constructed two general algorithms on the basis of random-scan Gibbs sampler, these are namely; an irreversible Gibbs sampler (IGS) and an irreversible Metropolized-Gibbs sampler (IMGS), both of which violate DBC but ensure invariance through SDBC. We test the algorithms on the 4-state Potts model and demonstrate numerically that both IGS and IMGS are not only superior to their respective reversible counter-parts which satisfy the strict DBC, but also outperform the generalized form of the IMH algorithm in reducing autocorrelation times.

\section*{Detailed balance condition} \label{section DBC}

In this paper we mostly consider a physical system with discrete state space $\Omega = \lbrace 1,..., S\rbrace$ where $S$ is the total number of states. We wish to sample from a target probability distribution $\bm{\pi} = \left(\pi_1,...,\pi_{S}\right)$ with $\pi_i>0$ and $\sum_{i = 1}^{S}\pi_i = 1$. We therefore use an MCMC algorithm to construct a Markov chain requiring that the stationary distribution of the chain coincide with the invariant target distribution $\bm{\pi}$. To do this the transition matrix $\bm{T} = \left(T_{ij}\right)_{i,j \in \Omega}$ of the Markov chain must satisfy the balance condition (BC) given by 
\begin{equation}
\pi_i = \sum_{j}\pi_jT_{ji} \,\,\,\,\,\, \forall \, i.
\end{equation}

\begin{figure}[htp]
\centering
\includegraphics[width=.3\textwidth ]{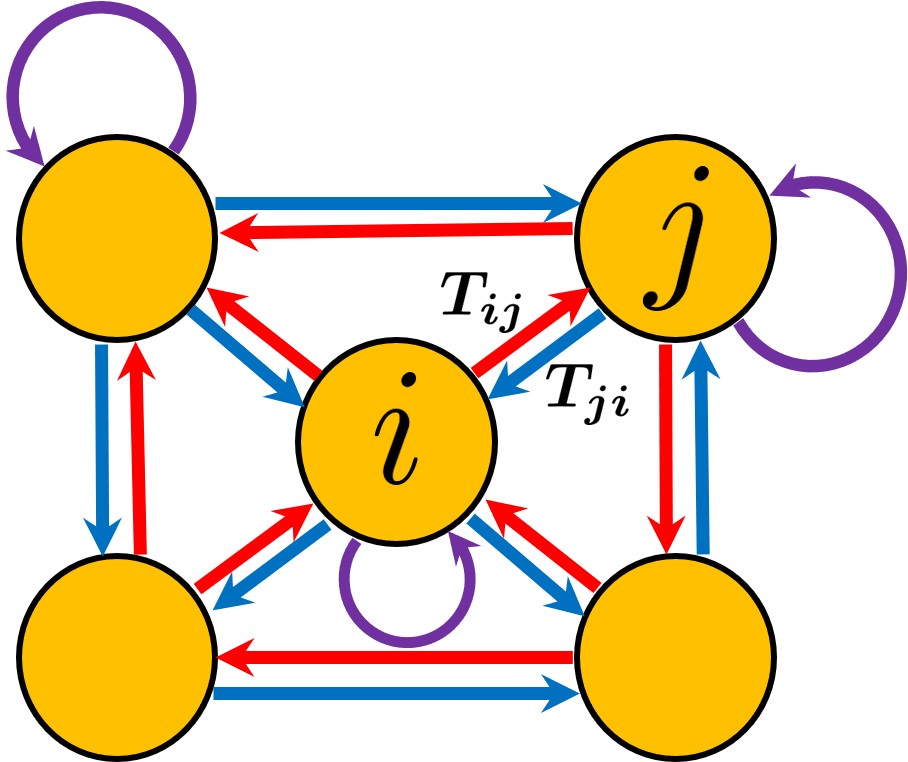}
\caption{The transition matrix $\bm{T}$ represented here schematically for five discrete states whereby the transition probability from one state to another is indicated by a single arrow.}
\label{graphical T matrix}
\end{figure}

The transition matrix must also meet the ergodicity requirement \cite{krauth book}. In the construction of MCMC algorithms the detailed balance condition (DBC),
\begin{equation}\label{DBC condition}
\pi_iT_{ij} = \pi_jT_{ji} ,
\end{equation}
has been widely imposed upon the transition matrix as a sufficient condition for satisfying BC, such Markov chains with DBC are commonly referred to as reversible Markov chains while those not meeting DBC are irreversible Markov chains.
   
\subsection*{Metropolis-Hastings algorithm}

The Metropolis-Hastings algorithm \cite{Metropolis}, arguably the most commonly used MCMC algorithm, enforces the detailed balance condition by requiring that the stochastic flow $v_{ij} = \pi_iT_{ij}$ is balanced out by its inverse flow $v_{ji} = \pi_jT_{ji}$. The transition matrix $T_{ij}$ can be written as 
\begin{align}\label{metropolis transition prob}
T_{ij} &= Q_{ij}A_{ij} \,\,\,\,\,\, \forall j \neq i,\\
T_{ii} &= 1 - \sum_{j\neq i} T_{ij}, \nonumber
\end{align} 
where  $\bm{Q} = \left(Q_{ij}\right)_{i,j \in \Omega}$ and $\bm{A} = \left(A_{ij}\right)_{i,j \in \Omega}$ are $S\times S$ matrices whose elements denote the proposal and acceptance probabilities respectively. Letting $X^{(t)}$ to denote the state of the system in $\Omega$ after $t$ iterations the general execution of the Metropolis-Hastings algorithm is then given in Algorithm.\ref{MH Algorithm}.

\begin{algorithm}[H]
  \begin{algorithmic}[1]
    \Input{Initialize $X^{(0)} = i \in \Omega$}
       \State \textbf{For} $t = 0,...,T-1$
          \State Pick a new candidate state $j \neq i \in \Omega$ with the probability $Q_{ij}$.
          \State Accept the new state $X^{(t+1)} = j$ with the probability $A_{ij}$.
          \State If the new state is rejected, assign $X^{(t+1)} = X^{(t)}$.
       \State \textbf{end for}
  \end{algorithmic}
  \caption{Metropolis-Hastings algorithm (MH)}
  \label{MH Algorithm}
\end{algorithm}

The original Metropolis algorithm \cite{Metropolis} assumed a symmetric proposal matrix $\bm{Q}$, it was later demonstrated by Hastings \cite{Hastings} that the proposal matrix need not be symmetric. The general form of the MH acceptance probability is therefore given by
\begin{equation} \label{metropolis acceptance prob}
A^{(\text{MH})}_{ij} = \text{min}\left(1, \frac{Q_{ji}\pi_j}{Q_{ij}\pi_i}\right).
\end{equation}
It is a simple exercise to demonstrate that the Metropolis-Hastings acceptance probability given in \eqref{metropolis acceptance prob} readily satisfies the balance condition through DBC given in \eqref{DBC condition}. The MH transition matrix,
\begin{equation}\label{metropolis transition matrix1}
T^{\text{(MH)}}_{ij} = Q_{ij}A^{(\text{MH})}_{ij},
\end{equation}   
therefore ensures the invariance of the target distribution $\bm{\pi}$.

\subsection*{Gibbs sampler}

Consider a general system with $N$ individual components whereby the state variable of the system $\bm{\sigma} \in \Omega$ is defined by a state vector $\bm{\sigma} = \left(\sigma_1,... ,\sigma_N\right)$ in the discrete state space $\Omega = \lbrace 1,...,S \rbrace$ with  $\sigma_k \in \lbrace 1,...,q \rbrace$ for $k = 1,...,N$. The state space therefore consists of $S = q^{N}$ number of configurations and the target distribution is $\bm{\pi}$. The Gibbs sampler (GS) \cite{Gibbs sampler}, also known as the Heat bath algorithm in statistical physics, updates only one component of the state vector, say $\sigma_k$, at a time. This component is assigned a new value sampled from its conditional distribution $\pi\left( \, \cdot \, \vert \bm{\sigma}_{-k}\right)$ where $\bm{\sigma}_{-k} = \left(\sigma_1, ...,\sigma_{k-1}, \sigma_{k+1}, ...,\sigma_N\right)$ are considered fixed. For convenience we let the state vector $\bm{\sigma}_k^{\nu} = (\sigma_k^{\nu}, \bm{\sigma}_{-k})$ to denote the state of the system where component $k$ is in state $\nu \in \lbrace 1,...,q \rbrace$ and the rest of the system is in state $\bm{\sigma}_{-k}$. A general execution of the random scan Gibbs sampler, whereby at each successive step a component of the system is selected to update uniformly at random, is given in Algorithm.\ref{Gibbs sampler algo}.

\begin{algorithm}[H]
  \begin{algorithmic}[1]
    \Input{Initialize $\bm{\sigma}^{(0)}  = \left(\sigma_1^{(0)}, ...,\sigma^{(0)}_N\right).$}
       \State \textbf{For} $t = 0,...,T-1$
          \State Pick a component $k \in \lbrace 1,...,N \rbrace$ uniformly at random.
          \State Draw a sample $\sigma^{(t+1)}_k \sim \pi\left(\, \cdot \,\,\, \vert \bm{\sigma}^{(t)}_{-k}\right)$.
          \State Assign $\bm{\sigma}^{(t+1)} = \left(\sigma^{(t)}_1, ..., \sigma^{(t)}_{k-1}, \sigma^{(t+1)}_k, \sigma^{(t)}_{k+1}, ..., \sigma^{(t)}_{N}\right).$
       \State \textbf{end for}
  \end{algorithmic}
  \caption{Gibbs sampler (GS)}
  \label{Gibbs sampler algo}
\end{algorithm}

The Gibbs sampler is a special case of the Metropolis-Hastings criteria whereby every proposal is accepted. For the random scan Gibbs sampler this can be easily demonstrated by letting the proposal $Q\left(\sigma_k', \bm{\sigma}_{-k}\vert \sigma_k, \bm{\sigma}_{-k} \right) = \frac{1}{N}\pi\left(\sigma_k' \vert \bm{\sigma}_{-k}\right)$ for $\sigma_k' \in \lbrace 1,...,q \rbrace$, and the acceptance $A\left(\sigma_k', \bm{\sigma}_{-k}\vert \sigma_k, \bm{\sigma}_{-k} \right) = \text{min}\left(1, r\right)$, where the ratio $r$ may then be written as
\begin{align}
r &= \frac{Q\left(\sigma_k, \bm{\sigma}_{-k}\vert \sigma_k', \bm{\sigma}_{-k} \right)\pi\left(\sigma_k',\bm{\sigma}_{-k}\right)}{Q\left(\sigma_k', \bm{\sigma}_{-k}\vert \sigma_k, \bm{\sigma}_{-k} \right) \pi\left(\sigma_k,\bm{\sigma}_{-k}\right)} \nonumber \\
   &= \frac{\pi\left(\sigma_k \vert \bm{\sigma}_{-k}\right)\pi\left(\sigma_k',\bm{\sigma}_{-k}\right)}{\pi\left(\sigma_k' \vert \bm{\sigma}_{-k}\right)\pi\left(\sigma_k,\bm{\sigma}_{-k}\right)} \nonumber \\
   & = 1.
\end{align}
The acceptance probability of each proposal is therefore exactly 1. As a special case of Metropolis-Hastings criteria the Gibbs sampler readily ensures the invariance of the target distribution $\bm{\pi}$. The random scan Gibbs sampler given in Algorithm.\ref{Gibbs sampler algo} satisfies DBC, in practice however the Gibbs sampling updates are commonly applied to each system component in sequence which produces a non-reversible chain. In general terminology, in the sequential updating scheme \cite{Ren} each component $k$ of the system has an associated transition matrix $\bm{C}^{(k)}$, and is for example updated with Metropolis-Hastings acceptance or a new value sampled from its conditional distribution as in the Gibbs sampler. Therefore $\bm{C}^{(k)}$ satisfies DBC locally: $\pi(\bm{\sigma}_k^{\mu})C^{(k)}(\bm{\sigma}_k^{\nu} \vert \bm{\sigma}_k^{\mu}) = \pi(\bm{\sigma}_k^{\nu})C^{(k)}(\bm{\sigma}_k^{\mu} \vert \bm{\sigma}_k^{\nu})$ and by implication also the balance condition: $\bm{\pi} = \bm{\pi}\bm{C}^{(k)}$. On the other hand given that the updating sequence is fixed and iterates from component $1$ to $N$ sequentially, then the transition matrix $\bm{S} = \prod_{k = 1}^{N}\bm{C}^{(k)}$ of each full sweep (i.e $N$ trial moves) breaks DBC: $\pi(\bm{\sigma}_k^{\mu})S(\bm{\sigma}_k^{\nu} \vert \bm{\sigma}_k^{\mu}) \neq \pi(\bm{\sigma}_k^{\nu})S(\bm{\sigma}_k^{\mu} \vert \bm{\sigma}_k^{\nu})$ but ensures invariance by satisfying BC: $\bm{\pi}\bm{S} = \bm{\pi}\prod_{k = 1}^{N}\bm{C}^{(k)} = \bm{\pi}$. The transition matrix $\bm{S}$ breaks DBC because for a given component $k$ an immediate reversal of a Monte Carlo move is not possible within a sweep.

Given that a component $k \in \lbrace 1,...,N \rbrace$ is sampled, the Gibbs transition rate $G(\bm{\sigma}_{k}^{\nu} \vert \bm{\sigma}_k^{\mu})$ from state $\bm{\sigma}_k^{\mu}$ to $\bm{\sigma}_k^{\nu}$ is then simply the conditional distribution given $\bm{\sigma}_{-k}$:

\begin{equation}\label{Gibbs transition matrix}
G(\bm{\sigma}_{k}^{\nu} \vert \bm{\sigma}_k^{\mu}) =  \frac{\pi\left(\bm{\sigma}_k^{\nu}\right)}{\sum\limits_{l = 1}^{q}\pi\left(\bm{\sigma}_k^{l}\right)}  \,\,\,\,\,\, \forall \,  \nu \in \lbrace 1,...,q \rbrace.
\end{equation} 

\noindent Notice that the transition rate to a new value $\nu$ is independent of the initial value $\mu$. We also point out that for $q = 2$ in \eqref{Gibbs transition matrix} the Gibbs sampler is equivalent to Barker's method \cite{Barker}, also known as Glauber dynamics in physics \cite{Glauber Dynamics}. Peskun \cite{Peskun} has  shown that within DBC the Metropolis-Hastings criteria is superior to Barker's method as it provides a more efficient sampling of the state space by returning smaller probabilities of remaining in the current state. While the Gibbs sampler described here does not involve an accept-reject criteria, one may regard a move rejected if the new candidate state $\nu$ is the current state $\mu$.

\subsection*{Metropolized-Gibbs sampler}

In this paper we term a \emph{Metropolized-Gibbs} sampler (MGS) to refer to  Liu's modification \cite{Liu1, Liu2} of the discrete state, random scan Gibbs sampler which is shown to increase the probability of transition to all states $j \in \Omega $ except for the current state $i \in \Omega $. The random scan Gibbs sampler satisfies detailed balance, the Metropolized-Gibbs sampler is an improvement on the random scan Gibbs sampler motivated directly by Peskun's theorem \cite{Peskun}: A Markov chain with smaller diagonal elements (i.e. smaller probability of remaining in the current state) provides a more efficient exploration of the state space and thus returns estimates with smaller asymptotic variance than a transition matrix with larger corresponding diagonal elements.  The modification on the random scan Gibbs sampler involves picking a component $k \in \lbrace 1,...,N \rbrace$ uniformly at random and excluding the current value $\sigma_k = \mu$ when proposing a new candidate value $\sigma_k = \nu$. The new candidate value $\sigma_k = \nu \neq \mu$ is now  proposed with the probability
\begin{equation}\label{MGS proposal prob}
Q(\bm{\sigma}_k^{\nu} \vert \bm{\sigma}_k^{\mu}) = \frac{G(\bm{\sigma}_{k}^{\nu} \vert \bm{\sigma}_k^{\mu})}{1 - G(\bm{\sigma}_{k}^{\mu} \vert \bm{\sigma}_k^{\nu})} \,\, \forall \,\, \nu \neq \mu.
\end{equation}
The Metropolis-Hastings acceptance probability \eqref{metropolis acceptance prob} for the state $\bm{\sigma}_k^{\nu}$ is then given by
\begin{equation}
A(\bm{\sigma}_k^{\nu} \vert \bm{\sigma}_k^{\mu}) = \text{min}\left[1, \frac{1 - G(\bm{\sigma}_k^{\mu} \vert \bm{\sigma}_k^{\nu})}{1 - G(\bm{\sigma}_k^{\nu} \vert \bm{\sigma}_k^{\mu})}\right]  \,\, \forall \,\, \nu \neq \mu,
\end{equation}
whereby upon rejection we retain the current state $\bm{\sigma}_k^{\mu}$. This gives a reversible transition matrix for the Metropolized-Gibbs sampler:
\begin{align}
M(\bm{\sigma}_k^{\nu} \vert \bm{\sigma}_k^{\mu}) &= \text{min}\left(\frac{G(\bm{\sigma}_k^{\nu} \vert \bm{\sigma}_k^{\mu})}{1 - G(\bm{\sigma}_k^{\mu} \vert \bm{\sigma}_k^{\nu})}, \frac{G(\bm{\sigma}_k^{\nu} \vert \bm{\sigma}_k^{\mu})}{1 - G(\bm{\sigma}_k^{\nu} \vert \bm{\sigma}_k^{\mu})}\right)  \,\,\,\,\, \forall \,\, \nu \neq \mu,  \label{MG transition rate}\\
M(\bm{\sigma}_k^{\mu} \vert \bm{\sigma}_k^{\mu}) &= 1 - \sum\limits_{\nu \neq \mu}M(\bm{\sigma}_k^{\nu} \vert \bm{\sigma}_k^{\mu}), \nonumber
\end{align}
which readily satisfies DBC. Note that when the denominator in \eqref{MGS proposal prob} vanishes, the transition matrix element in \eqref{MG transition rate} also vanishes, resulting in the rejection of the move. For practical implementation of the algorithm it is therefore recommended to make direct use of the transition matrix in (10). The optimality of MGS over the random scan Gibbs sampler follows from the same argument Peskun \cite{Peskun} put forward to show the superiority of Metropolis-Hastings criteria over other methods for swaps between two states: by excluding the current state when proposing a new candidate state the MGS updates tend to drive the Markov chain away from the current state. This may be further appreciated by noting that for $q = 2$ the MGS decomposes to the Metropolis-Hastings criteria whereas the standard Gibbs sampler becomes equivalent to Barker's criteria, a criteria shown to be less efficient than Metropolis-Hastings within DBC \cite{Peskun}. Furthermore we point out that just as in the Gibbs sampling updates, the MGS sampling updates too can be applied to each system component $k \in \lbrace 1,...,N \rbrace$ in sequence, in which case DBC is satisfied only locally. Equations \eqref{Gibbs transition matrix} and \eqref{MG transition rate} are thus valid regardless of how the system component $k$ is picked from the set $\lbrace 1,...,N \rbrace$. 

\section*{Lifting and the skewed detailed balance condition}\label{section SDBC}

In the lifting framework of Diaconis et al.\cite{Diaconis} the state space and the target distribution are extended by creating a duplicate replica of the system, each replica characterised by a lifting variable, and each state in the state space therefore acquiring two copies, one in each replica. An irreversible lifted Markov chain is thus propagated in this enlarged state space by a transition matrix that violates DBC but ensures invariance of the target distribution by satisfying BC. We provide in this section a brief review of the lifting framework with skewed detailed balance condition to construct irreversible Markov chains, as proposed by Turitsyn et al. \cite{Turitsyn} and extensively studied by Sakai and Hukushima \cite{Sakai Hukushima 1D, Sakai Hukushima 2D, Sakai Hukushima eigenvalue, Sakai Hukushima simulated tempering}.

We introduce an auxiliary or lifting variable $\varepsilon \in \lbrace +1,-1 \rbrace$ and effectively double the state space $\Omega$ so that the extended state space $\widetilde{\Omega}: = \Omega \times \lbrace +, - \rbrace$ consists of two replicas marked by $\varepsilon = \pm$. In this light the extended target distribution $\widetilde{\bm{\pi}}$ is given by 
\begin{align}\label{extended target dist}
\widetilde{\bm{\pi}} &= \left(\pi_{(1,+)},..., \pi_{(S,+)}, \pi_{(1,-)},..., \pi_{(S,-)}\right)\nonumber \\
            &= \frac{1}{2}\left(\bm{\pi},\bm{\pi}\right),
\end{align}

\begin{figure}[t!]
\centering
\includegraphics[width=0.45\textwidth]{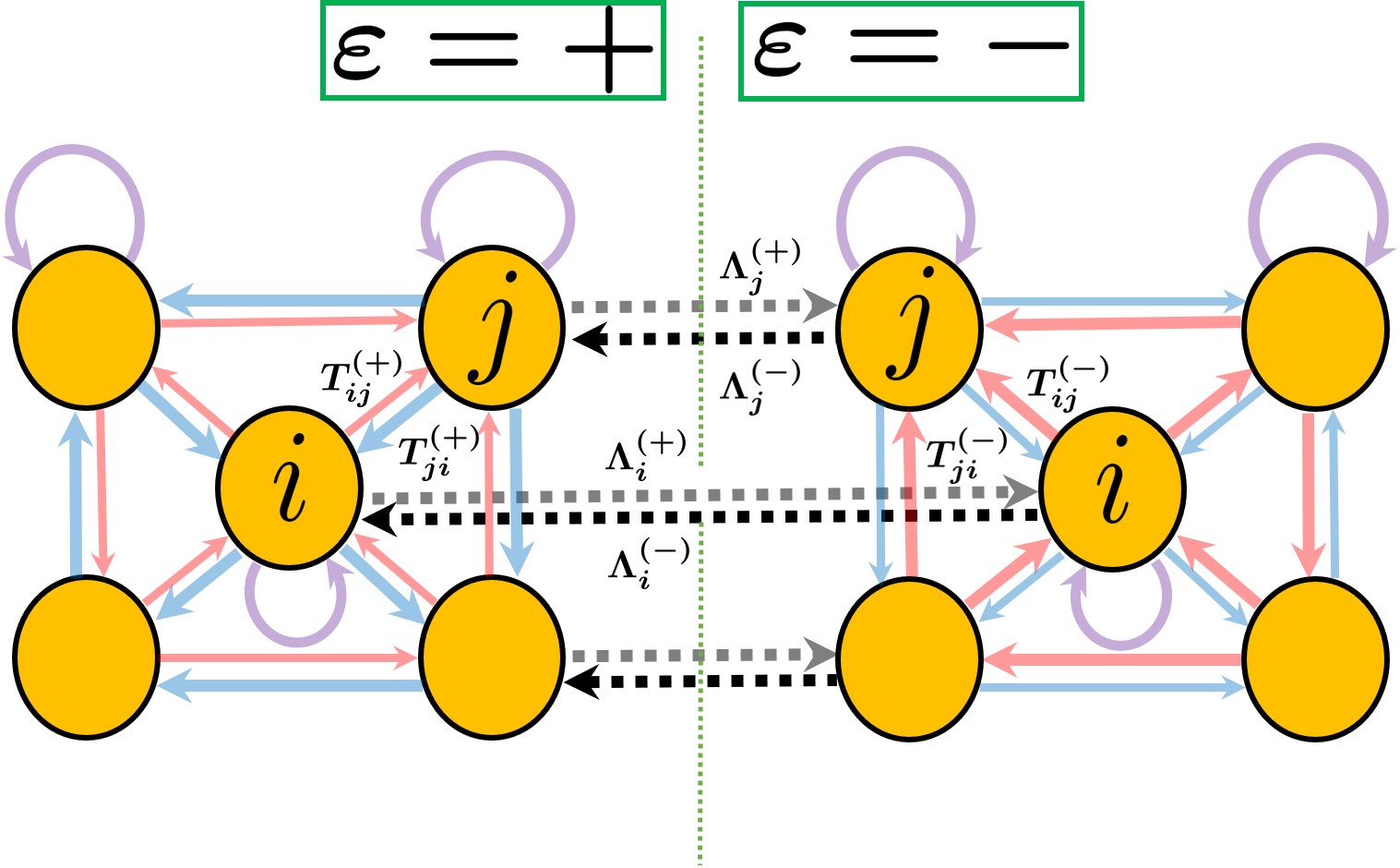}
\caption{Graphical representation of the Markov chain transition matrix $\widetilde{\bm{T}}$ on extended state space. The positive and negative  replicas are indicated by $\varepsilon = \pm$ left and right of the vertical dashed line respectively. In addition to \textit{intra-replica} transition flows $T^{(\pm)}_{ij}$ within states  indicated by solid arrows, we now have \textit{inter-replica} transition flows $\Lambda^{(\pm)}$ indicated by dashed arrows, which effectively execute the lifting mechanism.}
\label{graphical extended T matrix}
\end{figure}

\noindent where $\pi_{(i,\varepsilon)}$ concerns the probability of the state $(i,\varepsilon)$. The extended target distribution $\widetilde{\bm{\pi}}$ is independent of the auxiliary variable $\varepsilon$ so that 
\begin{equation}
\widetilde{\pi}_{(i,\varepsilon)} = \widetilde{\pi}_{(i,-\varepsilon)} 
\end{equation}
It is straightforward to show that the expectation value $\text{E}_{\widetilde{\bm{\pi}}}\left[f\right]$ of an observable $f$ with respect to the extended target distribution $\widetilde{\bm{\pi}}$ remains unchanged from that with respect to the original distribution $\bm{\pi}$, i.e.
\begin{align}\label{extended expectation}
\text{E}_{\widetilde{\bm{\pi}}}\left[f\right] &= \sum_{\varepsilon = \pm}\sum_{i = 1}^{S}\widetilde{\pi}_{(i,\varepsilon)}f_{(i,\varepsilon)}\nonumber\\
                                         & = \sum_{\varepsilon = \pm}\sum_{i = 1}^{S}\frac{\pi_{i}}{2}f_{(i,\varepsilon)}\nonumber\\
                                         & = \sum_{i = 1}^{S}\pi_{i}f_{i}\nonumber\\
                                         & = \text{E}_{\bm{\pi}}\left[f\right],
\end{align} 
where $f_{(i,\varepsilon)}$ denotes the realisation of the observable $f$ at state $(i,\varepsilon)$ and we have assumed $f_{(i,\varepsilon
)}  = f_{(i,-\varepsilon)} = f_{i}$ so that the observable $f$ is independent of $\varepsilon$. 

The transition matrix $\widetilde{\bm{T}}$ of the Markov chain on extended space $\widetilde{\Omega}$ is given by
\begin{equation}
 \widetilde{\bm{T}}=
  \left( {\begin{array}{cc}
   T^{(+)} & \Lambda^{(+)} \\
   \Lambda^{(-)} & T^{(-)} \\
  \end{array} } \right),
\end{equation}
where $T^{(\pm)} = (T^{(\pm)}_{ij})_{ij \in \Omega}\geq 0$ indicates the \textit{intra-replica} transition probability from state $i$ to $j$ in respective $\varepsilon = \pm$ replicas. The positive and diagonal \textit{inter-replica} matrices $\Lambda^{(\pm)} = \text{diag}(\Lambda_i^{(\pm)})_{i \in \Omega}\geq 0$ denotes the transition probability from state $(i,\varepsilon)$ to $(i,-\varepsilon)$ as shown in Fig.(\ref{graphical extended T matrix}). 

Normalization of probability is now explicitly in the form
\begin{equation}
\sum_{j \in \Omega}T_{ij}^{(\varepsilon)} + \Lambda_i^{(\varepsilon)} = 1,\,\,\,\, \forall \,\, i \in \Omega.
\end{equation}
Assuming that $\widetilde{\bm{T}}$ is ergodic, the balance condition
\begin{equation}\label{BC condition revised}
\widetilde{\pi}_m = \sum_{n}\widetilde{\pi}_n\widetilde{T}_{nm}, \,\,\,\, \forall \,\, m,
\end{equation}
will then ensure that the stationary distribution of the transition matrix $\widetilde{\bm{T}}$ is the extended target distribution $\widetilde{\bm{\pi}}$.

The balance condition for extended transition matrix $\widetilde{\bm{T}}$ may explicitly be written as 
\begin{equation}\label{BC explicit}
\sum_{j \in \Omega} \pi_i T_{ij}^{(\varepsilon)} + \pi_i\Lambda_i^{(\varepsilon)} = \sum_{j \in \Omega} \pi_j T_{ji}^{(\varepsilon)} + \pi_i\Lambda_i^{(-\varepsilon)},\,\,\,\, \forall \,\, i \in \Omega,
\end{equation}
where we have made use of \eqref{extended target dist}. The balance condition in \eqref{BC explicit} can be satisfied by imposing SDBC, which is given by 
\begin{equation}\label{SDBC}
\pi_iT_{ij}^{(\varepsilon)} = \pi_jT_{ji}^{(-\varepsilon)}.
\end{equation}
This allows us to construct an \textit{intra-replica} transition probability $T_{ij}^{(\varepsilon)}$ for an irreversible Markov chain. SDBC requires that the stochastic flow $v^{(\varepsilon)}_{ij} = \pi_iT_{ij}^{(\varepsilon)}$ in one replica is balanced out by the inverse flow $v^{(-\varepsilon)}_{ji} = \pi_jT_{ji}^{(-\varepsilon)}$ in the other replica. Note that SDBC readily  breaks detailed balance condition, i.e. $\pi_iT_{ij}^{(\varepsilon)} \neq \pi_jT_{ji}^{(\varepsilon)}$. Furthermore forcing SDBC provides a guideline for the construction of the \textit{inter-replica} transition probability $\Lambda_i^{(\varepsilon)}$, this becomes immediately obvious when we insert \eqref{SDBC} into \eqref{BC explicit} to obtain
\begin{equation}\label{lambda solution}
\Lambda_i^{(\varepsilon)} - \Lambda_i^{(-\varepsilon)} = \sum_{\substack{j \in \Omega \\ j\neq i}}\left(T_{ij}^{(-\varepsilon)} - T_{ij}^{(\varepsilon)}\right).
\end{equation} 
The solution to \eqref{lambda solution} is not unique, but there exist several choices.  Turitsyn et al. \cite{Turitsyn} had originally proposed the form:
\begin{equation}\label{TCV type}
\Lambda_i^{(\varepsilon)} = \text{max}\left[0, \sum_{\substack{j \in \Omega \\ j\neq i}}\left(T_{ij}^{(-\varepsilon)} - T_{ij}^{(\varepsilon)}\right) \right],
\end{equation}
which is known as Turitsyn-Chertkov-Vucelja (TCV) type. Several other choices have been proposed and studied by Sakai and Hukushima \cite{Sakai Hukushima 1D, Sakai Hukushima eigenvalue}, however the transition probability of TCV type has been shown numerically to provide the largest reduction in integrated autocorrelation times \cite{Sakai Hukushima 1D}. The following alternative choice known as the Sakai-Hukushima 1 type (SH1) has been studied analytically and numerically for the 1D Ising model \cite{Sakai Hukushima 1D}:
\begin{equation}
\Lambda_i^{(\varepsilon)} = \sum_{\substack{j \in \Omega \\ j\neq i}}T_{ij}^{(-\varepsilon)}.
\end{equation} 

\subsection*{Irreversible Metropolis-Hastings algorithm}\label{section IMH}
An irreversible Metropolis-Hastings algorithm (IMH) with skewed detailed balance condition was constructed for the mean-field Ising model by Turitsyn et al. \cite{Turitsyn}, this algorithm was later adapted to be applicable to more general systems with discreet degrees of freedom \cite{Sakai Hukushima eigenvalue}. In this section we are motivated to generalise the works of Sakai and Hukushima on 1D and 2D Ising model \cite{Sakai Hukushima 1D,Sakai Hukushima 2D}. We construct an irreversible Metropolis-Hastings algorithm to be applicable to classical spin systems in general. Our work specifically follows a prototype recipe provided by Sakai and Hukushima \cite{Sakai Hukushima eigenvalue} for constructing an \textit{intra-replica} transition matrix $(T_{ij}^{(\varepsilon)})_{i,j \in \Omega}$ that readily satisfies SDBC given in \eqref{SDBC}. This involves the modification of the transition matrix $\bm{T} = (T_{ij})_{i,j \in \Omega}$ which satisfies DBC: $\pi_iT_{ij} = \pi_jT_{ji}$, by a \textit{skewness} function: $[\Theta_{ij}^{(\varepsilon)}]_{i,j \in \Omega}$, so that 
\begin{equation}\label{transition recipe}
T_{ij}^{(\varepsilon)} = \Theta_{ij}^{(\varepsilon)}T_{ij},
\end{equation}
where the first requirement,
\begin{equation}\label{skewness condition 1}
0 \leq \Theta_{ij} \leq1,
\end{equation}
ensures that $T_{ij}^{(\varepsilon)}$ is a probability and the second requirement,
\begin{equation}\label{skewness condition 2}
\Theta_{ij}^{(\varepsilon)} = \Theta_{ji}^{(-\varepsilon)},
\end{equation}
guarantees that the transition matrix $T_{ij}^{(\varepsilon)}$ satisfies SDBC in \eqref{SDBC}. 

The \textit{skewness} function can be constructed to  directly utilize the physics of the system. Sakai and Hukushima \cite{Sakai Hukushima 1D} present a skewness function that introduces a bias in the way the magnetisation of the system is sampled in the Ising model. We build on their form and present a skewness function that is readily applicable to classical spin systems in general, such as the Potts model and the classical XY model, and can be readily adapted to use any observable of interest $f$ as the lifting coordinate.

\subsubsection*{Potts Model}

As an example of a classical spin system we focus on the Potts model on a lattice with $N$ sites, however the ideas in this section are equally applicable to classical continuous spin models. The Potts model is a generalisation of the Ising model \cite{MCMC physics2}, with the Hamiltonian defined as
\begin{equation}\label{potts model general hamiltonian}
H\left(\bm{\sigma}\right) = -\sum\limits_{\langle k,l \rangle}J_{kl} \, \delta\left(\sigma_k,\sigma_l\right),
\end{equation}
where $\delta(\cdot)$ is the Kronecker delta function and the notation $\langle k,l \rangle$ indicates that sites $k$ and $l$ are nearest neighbours on the lattice. $J_{kl}$ denotes the interaction strength between $\sigma_k$ and $\sigma_l$.  We have defined a given state of the Potts model (i.e. a given configuration) with the state vector $\bm{\sigma}= \left(\sigma_1,...,\sigma_N\right) {\in \Omega} $  in the discrete state space $\Omega = \lbrace 1, ..., S \rbrace$ with $\sigma_k \in \lbrace 1,...,q \rbrace$ for $k = 1,...,N$. The state space therefore consists of $S = q^{N}$ number of configurations. As before, we use $\bm{\sigma}_k^{\nu} = \left(\sigma_k^{\nu},\bm{\sigma}_{-k}\right)$ to denote a given configuration where the spin at site $k$ is in state $\nu \in \lbrace 1,...,q \rbrace$ and the rest of the system is in state $\bm{\sigma}_{-k} = \left(\sigma_1, ...,\sigma_{k-1}, \sigma_{k+1}, ...,\sigma_N\right)$.

We now wish to sample from the target distribution $\pi(\bm{\sigma})$ given by the Gibbs-Boltzmann distribution at a given inverse temperature $\beta$:
\begin{equation}\label{Gibbs Bolztmann distribution}
\pi\left(\bm{\sigma}\right) = \frac{1}{Z(\beta)}e^{-\beta H(\bm{\sigma})},
\end{equation}
where $Z\left(\beta\right) = \sum\limits_{\Omega}e^{-\beta H\left(\bm{\sigma}\right)}$ defines the partition function for a given inverse temperature.  

In the notation we have just introduced, the \textit{intra-replica} transition from state $(\bm{\sigma}_{k}^{\mu},\varepsilon)$ to $(\bm{\sigma}_{k}^{\nu}, \varepsilon)$ is indicated by $T(\bm{\sigma}_k^{\nu}, \varepsilon \vert \bm{\sigma}_k^{\mu}, \varepsilon)$ whereas $\Lambda(\bm{\sigma}_k^{\mu}, -\varepsilon \vert \bm{\sigma}_k^{\mu},\varepsilon,)$ indicates \textit{inter-replica} transition from state $(\bm{\sigma}_k^{\mu},\varepsilon)$ to $(\bm{\sigma}_k^{\mu},-\varepsilon)$. The balance condition in equation \eqref{BC explicit} may be expressed as 
\begin{align}
\sum\limits_{k, \nu}T(\bm{\sigma}_k^{\nu}, \varepsilon \vert \bm{\sigma}_k^{\mu}, \varepsilon)\widetilde{\pi}\left(\bm{\sigma}_k^{\mu},\varepsilon\right) &+ \Lambda\left(\bm{\sigma}_k^{\mu}, -\varepsilon \vert \bm{\sigma}_k^{\mu}, \varepsilon \right)\widetilde{\pi}\left(\bm{\sigma}_k^{\mu}, \varepsilon\right)\\
& = \sum\limits_{k, \nu}T(\bm{\sigma}_k^{\mu}, \varepsilon \vert \bm{\sigma}_k^{\nu}, \varepsilon )\widetilde{\pi}(\bm{\sigma}_k^{\nu}, \varepsilon) + \Lambda\left(\bm{\sigma}_k^{\mu}, \varepsilon \vert \bm{\sigma}_k^{\mu},-\varepsilon \right)\widetilde{\pi}\left(\bm{\sigma}_k^{\mu}, -\varepsilon \right),\nonumber
\end{align}
where the extended target distribution is given by \eqref{extended target dist}: $\widetilde{\pi}\left(\bm{\sigma}, \varepsilon \right) = \widetilde{\pi}\left(\bm{\sigma}, -\varepsilon \right) = \frac{1}{2}\pi(\bm{\sigma})$. An irreversible Markov chain can be constructed by imposing SDBC given in \eqref{SDBC}: 
\begin{equation} \label{SDBC potts}
\pi\left(\bm{\sigma}_k^{\mu}\right)T(\bm{\sigma}_k^{\nu},\varepsilon \vert \bm{\sigma}_k^{\mu}, \varepsilon) = \pi(\bm{\sigma}_k^{\nu})T(\bm{\sigma}_k^{\mu}, -\varepsilon \vert \bm{\sigma}_k^{\nu}, -\varepsilon).
\end{equation}

To proceed, we construct the transition rate $T(\bm{\sigma}_k^{\nu},\varepsilon \vert \bm{\sigma}_k^{\mu}, \varepsilon)$ according to \eqref{transition recipe}. An example of a skewness function $\Theta(\bm{\sigma}, \varepsilon)$ that readily satisfies requirement \eqref{skewness condition 2} has been studied by Sakai and Hukushima for 1D and 2D Ising models \cite{Sakai Hukushima 1D,Sakai Hukushima 2D}, this is of the form:
\begin{equation}\label{sakai hukushima form}
\Theta(\bm{\sigma},\varepsilon) = \varphi\left[1 - \delta \varepsilon \sigma_k\right],
\end{equation}
whereby setting the constant $\varphi = 1/(1 + \delta)$ and $\delta \in [0,1]$, not to be confused with the Kronecker delta function, ensures that the skewness function satisfies requirement \eqref{skewness condition 1}. While the form in \eqref{sakai hukushima form} seems specific to the Ising model, the following adaptation is applicable to classical spin systems in general:
\begin{equation}\label{skewness adapted}
\Theta\left(\bm{\sigma},\varepsilon \right) = \varphi \left[1 + \delta\varepsilon \Phi(f)\right],
\end{equation}
where the function $\Phi(f)$ is defined as
\begin{equation}\label{projection coordinate}
\Phi(f) = \text{sgn}\left[f(\bm{\sigma}_k^{\nu}) - f(\bm{\sigma}_k^{\mu})\right],
\end{equation} 
with $f$ denoting the lifting coordinate or the observable of interest and the sign function defined as
\[
    \text{sgn}(x)= 
\begin{cases}
    -1,& \text{if}\,\,\,\, x < 0,\\
     \,\,\,\,0,& \text{if}\,\,\,\, x = 0,\\
     +1,& \text{if}\,\,\,\,x > 0,
\end{cases}
\]
so that \eqref{skewness adapted} satisfies requirement \eqref{skewness condition 1}. One can simply recover the special form in \eqref{sakai hukushima form} by setting the lifting coordinate $f$ as the magnetisation of the system for the Ising model. The form in \eqref{skewness adapted} is not only applicable to classical spin systems in general but it also readily utilizes any observable of interest $f$ as the lifting coordinate. It is a simple exercise to confirm that the skewness function in \eqref{skewness adapted} meets the requirement in  \eqref{skewness condition 2}, thus guaranteeing that $T(\bm{\sigma}_k^{\nu}, \varepsilon \vert \bm{\sigma}_k^{\mu}, \varepsilon)$ satisfies SDBC. 

The \textit{intra-replica} transition rate $T\left(\bm{\sigma}_k^{\nu},\varepsilon \vert \bm{\sigma}_k^{\mu}, \varepsilon \right)$ with SDBC may now be expressed in the form:

\begin{equation}\label{extended transition general spins}
T(\bm{\sigma}_k^{\nu}, \varepsilon \vert \bm{\sigma}_k^{\mu}, \varepsilon) = \varphi \left[1 + \delta\varepsilon \Phi(f)\right] T(\bm{\sigma}_k^{\nu} \vert \bm{\sigma}_k^{\mu}),
\end{equation}

\noindent where $T(\bm{\sigma}_k^{\nu} \vert \bm{\sigma}_k^{\mu})$ is a transition rate from state $\bm{\sigma}_k^{\mu}$ to $\bm{\sigma}_k^{\nu}$ with DBC: $\pi(\bm{\sigma}_k^{\mu})T(\bm{\sigma}_k^{\nu} \vert \bm{\sigma}_k^{\mu}) = \pi(\bm{\sigma}_k^{\nu})T(\bm{\sigma}_k^{\mu} \vert \bm{\sigma}_k^{\nu})$. The parameter $\delta$ in the skewness function, which we will refer to as the \textit{deviation parameter}, just as in the original form, determines the extend to which DBC is violated; DBC is recovered in \eqref{extended transition general spins} with $\delta = 0$. 

\begin{figure*}[t!]
\centering
\includegraphics[width=0.7\textwidth]{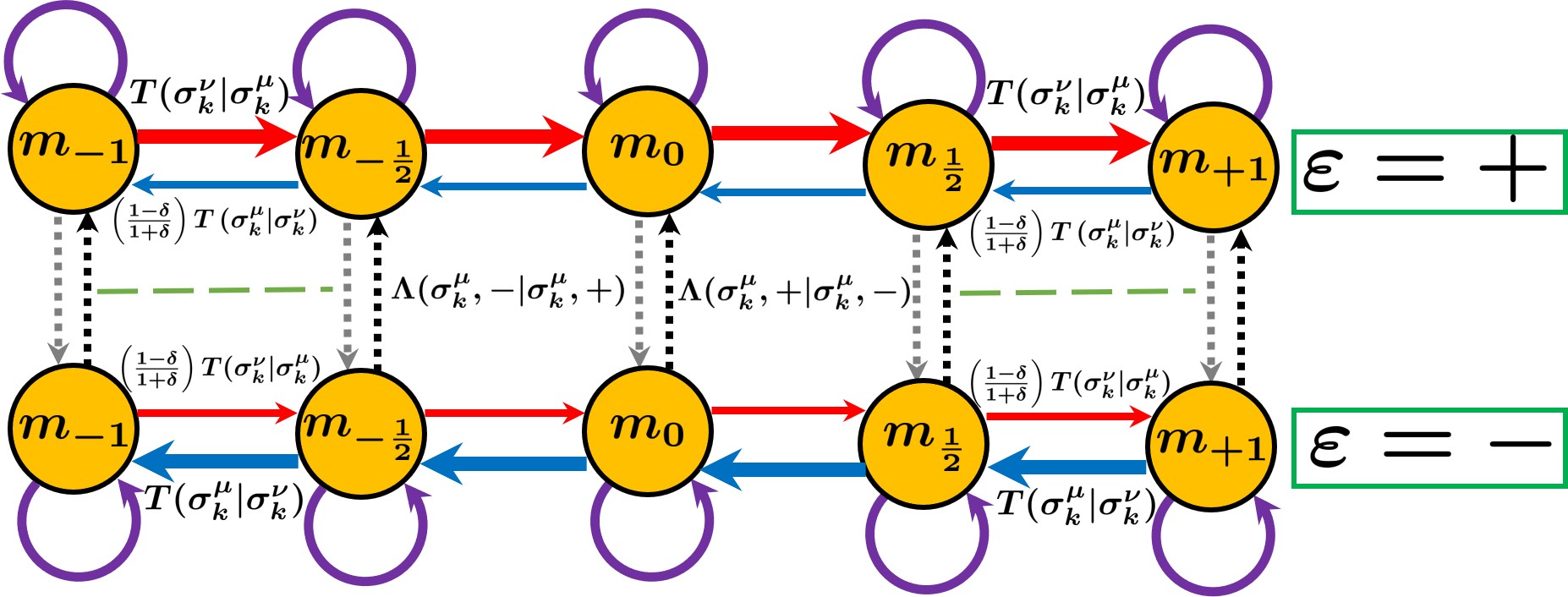}
\caption{Graphical representation of the transition probability $T\left(\bm{\sigma}_k^{\nu},\varepsilon \vert \bm{\sigma}_k^{\mu}, \varepsilon \right)$ for a simple $2 \times 2$ Ising system with magnetisation density $m$. The subscripts indicate the value of the magnetisation density and the solid arrows indicate $intra-replica$ transition flows between states, whereas $inter-replica$ lifting flows are indicated by dashed lines. We have assumed here a deviation parameter $\delta \neq 0$. In the positive replica $\varepsilon = +1$, moves that tend to increase the magnetisation of the system are favoured over those that tend to decrease the magnetisation, the opposite is true in the negative replica, $\varepsilon = -1$.}
\label{transition graph Ising}
\end{figure*}

The argument of the sign function in \eqref{projection coordinate}: $ \Delta f = f(\bm{\sigma}_k^{\nu}) - f(\bm{\sigma}_k^{\mu})$, simply denotes the change in the observable $f$ of the system if the spin at site $k$ acquires a new value $\nu$. To better understand how the transition rate in \eqref{extended transition general spins} introduces bias in the way the observable $f$ is sampled, it is helpful to consider two distinct scenarios: $\left( \varepsilon = \pm 1, \Phi(f) = \pm 1 \right)$ and $\left( \varepsilon = \pm 1, \Phi(f) = \mp 1 \right)$. The transition rate in \eqref{extended transition general spins} then decomposes to 

\[
    T\left(\bm{\sigma}_k^{\nu},\varepsilon \vert \bm{\sigma}_k^{\mu}, \varepsilon \right) = 
\begin{cases}
   \,\,\,\, T\left(\bm{\sigma}_k^{\nu} \vert \bm{\sigma}_k^{\mu}\right) & \text{for}\,\,\,\, \left( \varepsilon = \pm 1, \Phi(f) = \pm 1 \right), \\
     \,\,\,\, \left( \frac{1 - \delta}{1 + \delta} \right)T\left(\bm{\sigma}_k^{\nu} \vert \bm{\sigma}_k^{\mu}\right)& \text{for}\,\,\,\, \left( \varepsilon = \pm 1, \Phi(f) = \mp 1 \right).
\end{cases}
\]
A visual representation of the biased sampling imposed by the transition rate $T\left(\bm{\sigma}_k^{\nu},\varepsilon \vert \bm{\sigma}_k^{\mu}, \varepsilon \right)$  is shown in Fig.(\ref{transition graph Ising}) for a $2\times2$  Ising model with $N = 4$ sites, where we have chosen to set the projection coordinate $f$ as the magnetisation density of the system defined as 
\begin{equation}\label{magnetisation density}
m(\bm{\sigma}) = \frac{1}{N}\sum_{k = 1}^{N} \sigma_k.
\end{equation}

In Fig.(\ref{transition graph Ising}) we observe that for $\varepsilon = +1$ replica and $\delta \neq 0$ the transition rate $T\left(\bm{\sigma}_k^{\nu},\varepsilon \vert \bm{\sigma}_k^{\mu}, \varepsilon \right)$ is biased towards MCMC moves that tend to increase the magnetisation density $m$, whereas the transition rate of moves that propose to decrease $m$ are penalized with a factor $(1 -\delta)/(1 + \delta) < 1$; the opposite is true in $\varepsilon = -1$ replica. The selective sampling bias enforced by the transition rate $T\left(\bm{\sigma}_k^{\nu},\varepsilon \vert \bm{\sigma}_k^{\mu}, \varepsilon \right)$ may be characterised as the system acquiring momentum in a selected direction in state space to climb out of minimum energy states and thus explore the state space more efficiently. 

In Algorithm.\ref{IMH algo} the prototype of the irreversible Metropolis-Hastings algorithm (IMH) of Turitsyn et al. \cite{Turitsyn} for mean-field Ising model is now adapted for a classical spin system in general. We have used the notation $\widetilde{X}^{(t)}$ as a state of extended state space $\widetilde{\Omega}$ after $t$ iterations.

 \begin{algorithm}[H]
  \begin{algorithmic}[1]
    \Input{Initialize $\widetilde{X}^{(0)} = \left(\bm{\sigma}, \varepsilon \right).$}
       \State \textbf{For} $t = 0, ...,T-1$
          \State Select a site $k \in \lbrace 1,...,N \rbrace$ uniformly at random.
          \State Given that $\sigma_k = \mu$, propose a new spin value $\nu \neq \mu$ using the probability distribution $Q\left(\bm{\sigma}_k^{\nu} \vert \bm{\sigma}_k^{\mu}\right)$.
          \State Accept the new state $\widetilde{X}^{(t+1)} = \left(\bm{\sigma}_k^{\nu},\varepsilon \right)$ with the acceptance probability
          \begin{equation}
               A\left(\bm{\sigma}_k^{\nu}, \varepsilon \vert\bm{\sigma}_k^{\mu}, \varepsilon\right) = \varphi\left[1 + \delta \varepsilon \Phi(f)\right]A(\bm{\sigma}_k^{\nu}\vert \bm{\sigma}_k^{\mu})_{\text{MH}}.
          \end{equation} 
          \State If the proposed state is rejected, accept the state $\widetilde{X}^{(t+1)} = \left(\bm{\sigma}_k^{\mu},-\varepsilon \right)$ with the probability
          \begin{equation}\label{replica flip prob}
            P\left(\bm{\sigma}_k^{\mu}, -\varepsilon \vert \bm{\sigma}_k^{\mu},\varepsilon\right) = \frac{\Lambda\left(\bm{\sigma}_k^{\mu}, -\varepsilon \vert \bm{\sigma}_k^{\mu}, \varepsilon \right)}{1 - \sum\limits_{k'}\sum\limits_{\nu' \neq \mu'}T\left(\bm{\sigma}_{k'}^{\nu'}, \varepsilon \vert \bm{\sigma}_{k'}^{\mu'}, \varepsilon \right)}.        
          \end{equation}
          \State If this is also rejected then set $\widetilde{X}^{(t+1)} = \widetilde{X}^{(t)}$.
       \State \textbf{end for}
  \end{algorithmic}
  \caption{IMH for classical spin systems}
  \label{IMH algo}
\end{algorithm}

\noindent The Metropolis-Hastings transition rate $T(\bm{\sigma}_k^{\nu} \vert \bm{\sigma}_k^{\mu})_{\text{MH}}$ is decomposed into proposal $Q(\bm{\sigma}_k^{\nu} \vert \bm{\sigma}_k^{\mu})$ and acceptance $A(\bm{\sigma}_k^{\nu} \vert \bm{\sigma}_k^{\mu})_{\text{MH}}$:

\begin{equation}\label{MH transition rate}
T(\bm{\sigma}_k^{\nu} \vert \bm{\sigma}_k^{\mu})_{\text{MH}} = Q(\bm{\sigma}_k^{\nu} \vert \bm{\sigma}_k^{\mu})A(\bm{\sigma}_k^{\nu} \vert \bm{\sigma}_k^{\mu})_{\text{MH}}\,\, ,
\end{equation}
where the MH acceptance rate, written explicitly in this notation, is in the form:
\begin{equation}\label{MH acceptance rate}
A(\bm{\sigma}_k^{\nu} \vert \bm{\sigma}_k^{\mu})_{\text{MH}} =  \text{min}\left[1,\frac{Q(\bm{\sigma}_k^{\mu} \vert \bm{\sigma}_k^{\nu})\pi(\bm{\sigma}_k^{\nu})}{Q(\bm{\sigma}_k^{\nu} \vert\bm{\sigma}_k^{\mu})\pi(\bm{\sigma}_k^{\mu})}\right].
\end{equation}
The \textit{inter-replica} transition rate is chosen to be that of TCV type given in \eqref{TCV type}:

\begin{equation}
\Lambda(\bm{\sigma}_k^{\mu}, -\varepsilon \vert \bm{\sigma}_k^{\mu}, \varepsilon) = \text{max}\left[0, \sum\limits_{k'}\sum\limits_{\nu' \neq \mu'}\left(T(\bm{\sigma}_{k'}^{\nu'}, -\varepsilon \vert \bm{\sigma}_{k'}^{\mu'}, -\varepsilon ) - T(\bm{\sigma}_{k'}^{\nu'}, \varepsilon \vert \bm{\sigma}_{k'}^{\mu'}, \varepsilon)\right)\right],
\end{equation}
however alternative forms are given by Sakai and Hukushima \cite{Sakai Hukushima eigenvalue}.

In Algorithm.\ref{IMH algo}, unless otherwise specified one Monte Carlo step $t$ is taken to be one iteration of steps (2)-(6) and $T$ denotes the total number of Monte Carlo steps. To evaluate the probability in \eqref{replica flip prob} summation with respect to the number of sites, and a summation over $(q-1)$ spin states at each site is required. In practice the summation is computed at the initial configuration and from then on simply updated at each successful spin flip at step (4).

\subsection*{Markov chains with SDBC on the basis of Gibbs sampler}\label{IGS and IMGS}

In this section we demonstrate that a Markov chain with SDBC can be constructed on the basis of the Gibbs sampler. The prototype algorithm presented by Turitsyn et al.\cite{Turitsyn} for the mean field Ising model had been developed on the basis of Metropolis-Hastings criteria, and although a general formulation of the irreversible Metropolis-Hastings (IMH) was later presented by Sakai and Hukushima for discrete state systems \cite{Sakai Hukushima eigenvalue}, both of these efforts however have constructed the irreversible counterpart of the Metropolis-Hastings transition as given in \eqref{metropolis transition prob}. Here we develop irreversible Markov chains with SDBC on the basis of the Gibbs sampler and the \textit{Metropolized-Gibbs} sampler that break DBC but satisfy BC on the basis of SDBC. The algorithms are developed to be applicable to general discrete state systems.

\subsubsection*{Irreversible Gibbs sampler}

Let us again consider a general system with $N$ individual components whereby the state variable of the system is defined by the state vector $\bm{\sigma} = \left(\sigma_1,... ,\sigma_N\right) \in \Omega$ in the discrete state space $\Omega = \lbrace 1,...,S \rbrace$ with $\sigma_k \in \lbrace 1,...,q \rbrace$ for $k = 1,...,N$. The state space consists of $S = q^{N}$ number of configurations and the target distribution is $\bm{\pi}$. As before we denote a given state of the system $\bm{\sigma}_k^{\mu} = \left(\sigma_k^{\mu}, \bm{\sigma}_{-k}\right)$ to indicate that component $k$ is in state $\mu \in \lbrace 1,...,q \rbrace$ while the rest of the system is in state $\bm{\sigma}_{-k} = \left(\sigma_1, ...,\sigma_{k-1}, \sigma_{k+1}, ...,\sigma_N\right)$. The Gibbs transition probability for component $k$ to acquire a new state $\nu \in \lbrace 1,...,q \rbrace$ is then given in \eqref{Gibbs transition matrix}.
The transition matrix $\bm{\mathcal{G}}$ for the irreversible Gibbs sampler (IGS) with SDBC can be constructed according to \eqref{transition recipe}: 
\begin{align}\label{IGS}
\mathcal{G}(\bm{\sigma}_k^{\nu}, \varepsilon \vert \bm{\sigma}_k^{\mu}, \varepsilon) &= \Theta(\bm{\sigma}, \varepsilon)G(\bm{\sigma}_k^{\nu} \vert \bm{\sigma}_k^{\mu})\,\,\,\,\,\ \forall \,\,\, \nu \neq \mu,  \\
\mathcal{G}(\bm{\sigma}_k^{\mu}, \varepsilon \vert \bm{\sigma}_k^{\mu}, \varepsilon) &= 1 - \sum\limits_{\nu \neq \mu}\mathcal{G}(\bm{\sigma}_k^{\nu}, \varepsilon \vert \bm{\sigma}_k^{\mu}, \varepsilon), \nonumber 
\end{align}
where the Gibbs transition $G(\bm{\sigma}_k^{\nu} \vert \bm{\sigma}_k^{\mu})$ is given in \eqref{Gibbs transition matrix} and the skewness function $\Theta(\bm{\sigma}, \varepsilon)$ meets requirement \eqref{skewness condition 1}. SDBC is readily satisfied by imposing condition \eqref{skewness condition 2} on the skewness function. The transition matrix in \eqref{IGS} will therefore propagate an irreversible Markov chain on the extended state space $\widetilde{\Omega}$, yet ensuring the invariance of the target distribution. 

In Algorithm.\ref{IGS algo} we demonstrate the execution of IGS for a general discrete state system where unless otherwise specified one Monte Carlo step $t$ is defined to be one iteration of steps (2)-(5) with $T$ denoting the total number of MC steps.

\begin{algorithm}[H]
  \begin{algorithmic}[1]
    \Input{Initialize $\widetilde{X}^{(0)} = \left(\bm{\sigma}, \varepsilon \right).$}
       \State \textbf{For} $t = 0, ...,T-1$
          \State Select a component $k \in \lbrace 1,...,N \rbrace$ uniformly at random.
          \State Supposing $\bm{\sigma} = \bm{\sigma}_k^{\mu}$, now assign $\widetilde{X}^{(t+1)} = \left(\bm{\sigma}_k^{\nu}, \varepsilon \right)$ with the  probability
          \begin{align}          
          \mathcal{G}(\bm{\sigma}_k^{\nu}, \varepsilon \vert \bm{\sigma}_k^{\mu}, \varepsilon) &= \Theta(\bm{\sigma}, \varepsilon)G(\bm{\sigma}_k^{\nu} \vert \bm{\sigma}_k^{\mu})\,\,\,\,\,\ \forall \,\,\, \nu \neq \mu, \\
\mathcal{G}(\bm{\sigma}_k^{\mu}, \varepsilon \vert \bm{\sigma}_k^{\mu}, \varepsilon) &= 1 - \sum\limits_{\nu \neq \mu}\mathcal{G}(\bm{\sigma}_k^{\nu}, \varepsilon \vert \bm{\sigma}_k^{\mu}, \varepsilon). \nonumber  
\end{align}

          \State If $\bm{\sigma}_k^{\nu} = \bm{\sigma}_k^{\mu}$, accept the state $\widetilde{X}^{(t+1)} = \left(\bm{\sigma}_k^{\mu},-\varepsilon \right)$ with the probability
          \begin{equation}\label{IGS epsilon flip prob}
            P\left(\bm{\sigma}_k^{\mu}, -\varepsilon \vert \bm{\sigma}_k^{\mu},\varepsilon\right) = \frac{\Lambda\left(\bm{\sigma}_k^{\mu}, -\varepsilon \vert \bm{\sigma}_k^{\mu}, \varepsilon \right)}{1 - \sum\limits_{k'}\sum\limits_{\nu' \neq \mu'}\mathcal{G}\left(\bm{\sigma}_{k'}^{\nu'}, \varepsilon \vert \bm{\sigma}_{k'}^{\mu'}, \varepsilon \right)}.          
          \end{equation}
          \State If this is also rejected then set $\widetilde{X}^{(t+1)} = \widetilde{X}^{(t)}$.
       \State \textbf{end for}
  \end{algorithmic}
  \caption{Irreversible Gibbs sampler (IGS)}
  \label{IGS algo}
\end{algorithm}

With the particular choice of the skewness function given in \eqref{skewness adapted} the irreversible Gibbs sampler can be readily applied to discrete state classical spin systems such as the Potts model. However we stress that with a careful construction of a skewness function, that utilizes the properties of the system in question, IGS is applicable to any system with discrete degrees of freedom. Considering a discrete state spin system with $N$ sites where $\sigma_k \in \lbrace 1,...,q \rbrace$ for $k = 1,...,N$, the evaluation of the probability in \eqref{IGS epsilon flip prob} now requires a summation over $(q-1)$ spin values at each site in addition to a summation over $N$ sites in the lattice. However we point out that the summation is in practice computed only once at the initial conditions and from then on simply updated at each successful spin-flip process, that is updated at step (3) where $\bm{\sigma}_k^{\nu} \neq \bm{\sigma}_k^{\mu}$. 

The \textit{inter-replica} transition rate of the TCV type is now of the form:

\begin{equation}
\Lambda(\bm{\sigma}_k^{\mu}, -\varepsilon \vert \bm{\sigma}_k^{\mu}, \varepsilon) = \text{max}\left[0, \sum\limits_{k'}\sum\limits_{\nu' \neq \mu'}\left(\mathcal{G}(\bm{\sigma}_{k'}^{\nu'}, -\varepsilon \vert \bm{\sigma}_{k'}^{\mu'}, -\varepsilon ) - \mathcal{G}(\bm{\sigma}_{k'}^{\nu'}, \varepsilon \vert \bm{\sigma}_{k'}^{\mu'}, \varepsilon)\right)\right].
\end{equation}
It is worth noting that for $q = 2$ (the Ising model), the IGS decomposes to the irreversible Glauber dynamics studied by Sakai and Hukushima \cite{Sakai Hukushima 1D,Sakai Hukushima 2D}.

\subsubsection*{Irreversible Metropolized-Gibbs sampler}

In this brief section we point out that an irreversible counter-part of the Metropolized-Gibbs sampler (MGS), which we will henceforth refer to as the irreversible Metropolized-Gibbs sampler (IMGS), can be constructed based on the SDBC. The construction of the corresponding transition matrix $\bm{\mathcal{M}}$ follows the same principle as that of IGS:
\begin{align}\label{IMGS}
\mathcal{M}(\bm{\sigma}_k^{\nu}, \varepsilon \vert \bm{\sigma}_k^{\mu}, \varepsilon) &= \Theta(\bm{\sigma}, \varepsilon)M(\bm{\sigma}_k^{\nu} \vert \bm{\sigma}_k^{\mu}) \,\,\,\,\,\ \forall \,\,\, \nu \neq \mu, \\
\mathcal{M}(\bm{\sigma}_k^{\mu}, \varepsilon \vert \bm{\sigma}_k^{\mu}, \varepsilon) &= 1 - \sum\limits_{\nu \neq \mu}\mathcal{M}(\bm{\sigma}_k^{\nu}, \varepsilon \vert \bm{\sigma}_k^{\mu}, \varepsilon), \nonumber 
\end{align}
where $M(\bm{\sigma}_k^{\nu} \vert \bm{\sigma}_k^{\mu})$ is the MGS transition matrix given in \eqref{MG transition rate}. The general execution of the algorithm follows the same steps as in Algorithm.\ref{IGS algo} except for the use of \eqref{IMGS} in steps (3) and (4). IMGS is equivalently applicable to general discrete state systems. Note that for a special case of $q = 2$ (the Ising model) IMGS and IMH, as given in Algorithm.\ref{IMH algo}, are equivalent. This should be obvious since the Metropolized-Gibbs transition given in \eqref{MG transition rate} is essentially the Metropolis-Hastings criteria for $q = 2$. The development of IMGS is directly motivated to check if the efficiency of the MGS over random scan Gibbs sampler \cite{Liu2} is replicated in their irreversible counter-parts with SDBC. 

\section*{MCMC simulations}\label{section simulation}

\subsection*{Performance analysis on 1D Potts model}

As an application of IMH, IGS and IMGS algorithms, we consider the 1-Dimensional $q = 4$ state Potts model with $N$ sites and first nearest neighbour interactions. The Hamiltonian of the system is then directly deduced from the general form given in \eqref{potts model general hamiltonian}:
\begin{equation}
H\left(\bm{\sigma}\right) = -\sum\limits_{k = 1}^{N}J_{k,k+1} \, \delta\left(\sigma_k, \sigma_{k+1}\right),
\end{equation}
where a periodic boundary condition $\sigma_{N+1} = \sigma_1$ is imposed and the interaction strength are all set to 1 so that $J_{k,k+1} = J = 1$ for $ k = 1,...,N$. We remind the reader that according to \eqref{extended expectation} the expectation value $\text{E}_{\widetilde{\bm{\pi}}}\left[f\right]$ of an observable $f = f\left(\bm{\sigma}, \varepsilon \right)$ with respect to the extended target distribution $\widetilde{\pi}\left(\bm{\sigma}, \varepsilon \right)$ remains unchanged from that with respect to the original distribution $\pi\left(\bm{\sigma}\right)$, i.e. $\text{E}_{\widetilde{\bm{\pi}}}\left[f\right] = \text{E}_{\bm{\pi}}\left[f\right]$. The expectation value $\text{E}_{\bm{\pi}}[f]$ over the equilibrium distribution $\pi\left(\bm{\sigma}\right)$ is then given by
\begin{equation}
\text{E}_{\bm{\pi}}[f] = \sum\limits_{\Omega}f(\bm{\sigma})\pi\left(\bm{\sigma}\right)
\end{equation}
where $\sum_{\Omega}$ indicates a sum over $S = q^{N}$ spin configurations. The equilibrium distribution $\pi(\bm{\sigma})$ is the Gibbs-Boltzmann distribution given in \eqref{Gibbs Bolztmann distribution} where we define  the inverse temperature $\beta = 1/\mathcal{T}$ in units where the Boltzmann constant $k_B$ is set to 1. 

In classical 1D systems the non-existence of phase-transition at any physically accessible temperature $\mathcal{T}$ has been presented in various arguments and theorems \cite{phase transition 2, phase transition 3, phase transition 4, phase transition 5}, a 1D potts model therefore exhibits no spontaneous magnetisation at any finite temperature. For the 1D Potts model under our consideration we have imposed periodic boundary conditions and have let all sites to be equivalent, so that $J_{k,k+1} = J$ for $k = 1,...,N$. The expectation value of the magnetisation density over the equilibrium distribution $\pi(\bm{\sigma})$ is then given by
\begin{equation}
\text{E}_{\bm{\pi}}\left[m\right] = \frac{1}{q}\sum\limits_{\sigma = 1}^{q}\sigma,
\end{equation}
where $\text{E}_{\bm{\pi}}[m] = 2.5$ for $q = 4$. 

For the simulations that follow we define the ensemble average $\langle f(t) \rangle$ at time $t$  of an observable $f = f(\bm{\sigma}, \varepsilon)$ as 
\begin{equation}\label{ensemble average}
\langle f(t) \rangle = \frac{1}{N_{\text{sim}}}\sum\limits_{i = 1}^{N_{\text{sim}}}f\left(\bm{\sigma}^i(t), \varepsilon^i(t)\right),
\end{equation}
where time is measured in number of MC-steps starting from the initial conditions. $N_{\text{sim}}$ denotes the number of independent simulated trajectories and  $f\left(\bm{\sigma}^i(t), \varepsilon^i(t)\right)$ the realisation of observable $f$ at time $t$ for trajectory $i$.

The integrated autocorrelation time $\tau_{int,f}$ for an observable $f$ is defined as
\begin{equation}
\tau_{int,f} = 1 + 2\sum\limits_{t = 1}^{\infty}C_f(t),
\end{equation}
where $C_f(t)$ denotes the autocorrelation function  given the measurements, $f_1,f_2,...,f_M$:
\begin{equation}
C_f(t) = \frac{\text{E}_{\bm{\pi}}[f(t' + t)f(t')] - \text{E}_{\bm{\pi}}[f(t')]^2}{\text{E}_{\bm{\pi}}[f^2(t')] - \text{E}_{\bm{\pi}}[f(t')]^2},
\end{equation}
with $t'$ set sufficiently large for equilibration when estimating $C_f(t)$. $\tau_{int,f}$ is commonly estimated through the relation 
\begin{equation}
\tau_{int, f} = \frac{\sigma^2_{f}}{\sigma^2_{0, f}},
\end{equation} 
where $\sigma^2_{0,f} = \text{E}_{\bm{\pi}}[f^2] - \text{E}_{\bm{\pi}}[f]^2$ indicates the variance for an independent sampling, i.e. the \textit{naive} variance of the raw time series data treated as though all the values were independently sampled. $\sigma^2_f$ is the asymptotic variance computed through batch means method using batch sizes much larger than $\tau_{int,f}$ \cite{Berg}. A large integrated autocorrelation time of observable $f$ therefore indicates a large corresponding asymptotic variance.

\subsubsection*{Magnetisation density as the lifting coordinate}

\begin{figure*}[t!]
\centering
\includegraphics[width=.33\textwidth, height=.25\textwidth,]{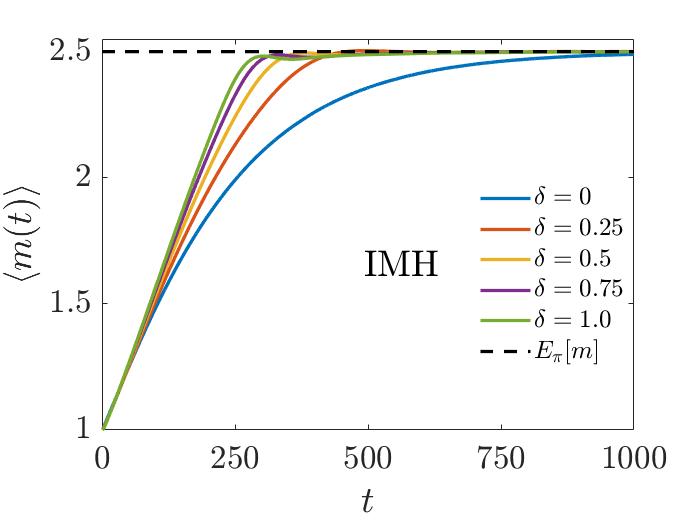}\hfill
\includegraphics[width=.33\textwidth, height=.25\textwidth,]{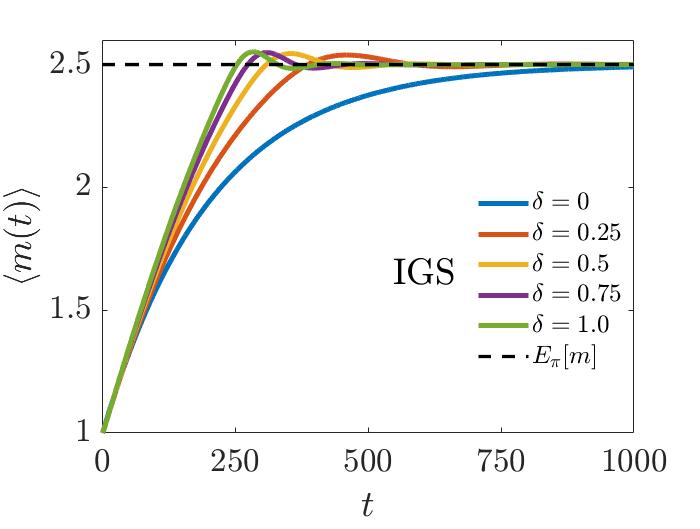}\hfill
\includegraphics[width=.33\textwidth, height=.25\textwidth,]{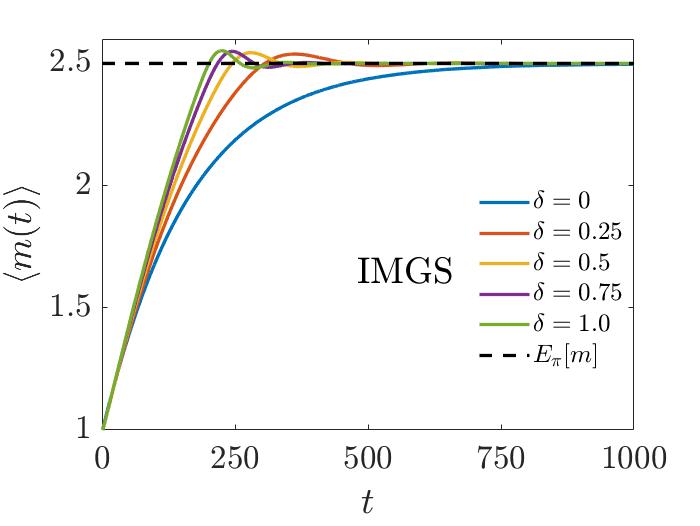}\hfill
\caption{\noindent The average trajectory tracing the evolution of magnetisation density with respect to time as in eq.\eqref{ensemble average}, starting from the initial conditions. The simulations were initialised with $\sigma_k = 1$ for $k = 1,...,N$, and a random assignment of $\varepsilon \in \lbrace +1,-1 \rbrace$. $N = 144$, temperature $\mathcal{T} = 2.0$, $N_{\text{sim}} = 10^5$ and $T = 2\times10^3$ MC-steps. The deviation parameter $\delta$ indicates deviation from the DBC.}
\label{magnetisation_vs_time}
\end{figure*}

We simulate the 1D 4-state Potts model with IMH, IGS and IMGS  whereby we deploy the skewness function introduced in \eqref{skewness adapted} and take the lifting coordinate $f$ to be the magnetisation density of the system. Fig.(\ref{magnetisation_vs_time}) shows the average trajectories tracing the evolution of the magnetisation density with respect to time. For all three algorithms it is observed that deviation from the DBC condition, $\delta = 0$, results in faster convergence to the equilibrium, which remains consistent with a similar study on 1D Ising model \cite{Sakai Hukushima 1D}. This speed-up in equilibration is likely attributed to the suppression of diffusive behaviour along key collective variables (reaction coordinates) as a result of breaking DBC, as a consequence the induction of probability flows in state space may accelerate exploration \cite{hot topic 7, Diaconis, Turitsyn}. 

\begin{figure*}[t!]
\centering
\includegraphics[width=.33\textwidth, height=.25\textwidth, ]{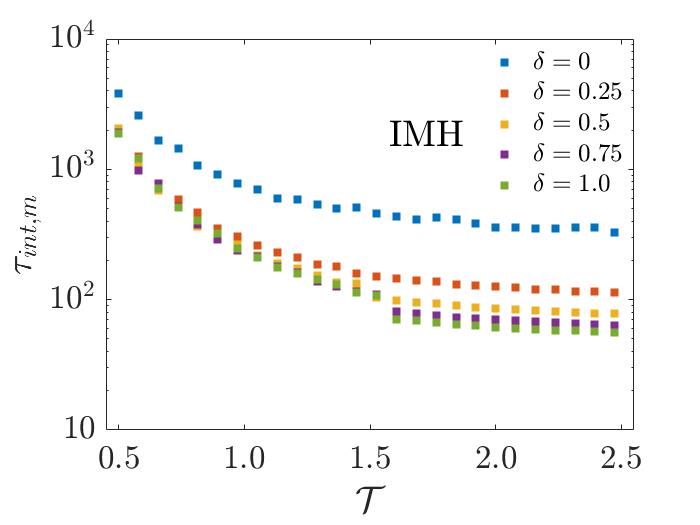}\hfill
\includegraphics[width=.33\textwidth, height=.25\textwidth,]{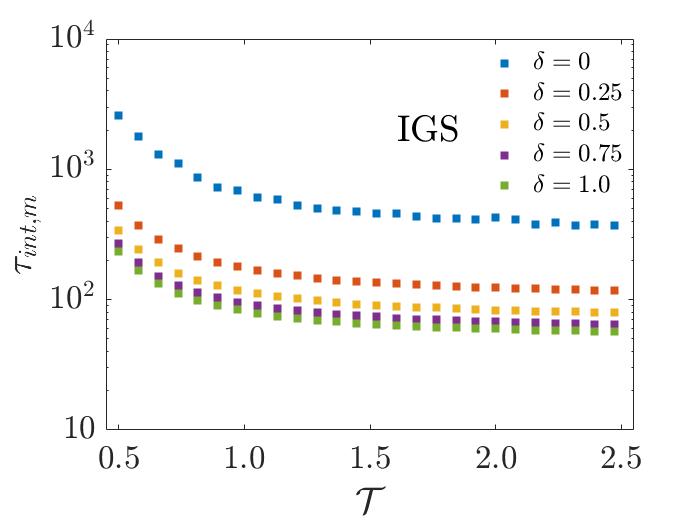}\hfill
\includegraphics[width=.33\textwidth, height=.25\textwidth,]{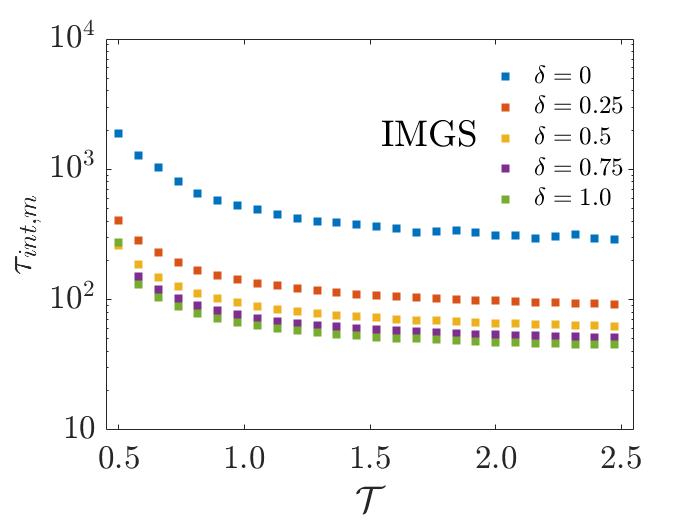}\hfill

\caption{Integrated autocorrelation times $\tau_{int,m}$ for the magnetisation density at 26 temperatures in the range $\mathcal{T} = 0.5-2.47$. The values are obtained from a very long single runs of the algorithms with $T = 10^7$ MC-steps and $N = 144$ sites. The deviation parameter $\delta$ indicates deviation from the DBC. }
\label{tau_vs_kT_magnetisation}
\end{figure*}

In Fig.(\ref{tau_vs_kT_magnetisation}) we show the integrated autocorrelation times of the magnetisation density $\tau_{int,m}$ for 26 temperatures in the range $\mathcal{T} = 0.5-2.47$, the values were obtained from a very long single runs of the algorithms. Deviation from the DBC condition, $\delta = 0$, induces a reduction in $\tau_{int,m}$ for all the temperatures in the given range, this observation is prevalent for all three algorithms albeit with varying degrees of reduction. Concerning the optimum deviation from the DBC condition, i.e. $\delta = 1$ for IMH, IGS and IMGS, we report that $\left[\tau_{(\delta = 0)}/\tau_{(\delta = 1)}\right]_{\text{IMH}} \sim 5.86$, $\left[\tau_{(\delta = 0)}/\tau_{(\delta = 1)}\right]_{\text{IGS}} \sim 7.12$ and $\left[\tau_{(\delta = 0)}/\tau_{(\delta = 1)}\right]_{\text{IMGS}} \sim 6.59$ at $\mathcal{T} = 2.0$ whereas $\left[\tau_{(\delta = 0)}/\tau_{(\delta = 1)}\right]_{\text{IMH}} \sim 2.33$, $\left[\tau_{(\delta = 0)}/\tau_{(\delta = 1)}\right]_{\text{IGS}} \sim 9.71$ and $\left[\tau_{(\delta = 0)}/\tau_{(\delta = 1)}\right]_{\text{IMGS}} \sim 9.93$ at $\mathcal{T} = 0.66$. At lower temperatures the reduction in $\tau_{int,m}$ (compared to their respective reversible counterparts) is evidently more profound for IGS and IMGS than that for the IMH. The IMGS in particular outperforms its reversible counterpart by almost an order of magnitude at $\mathcal{T} = 0.66$, compare this to a gain of only $\sim 2.33$ for IMH. 

\begin{figure*}[t!]
\centering
\includegraphics[width=.33\textwidth, height=.25\textwidth, ]{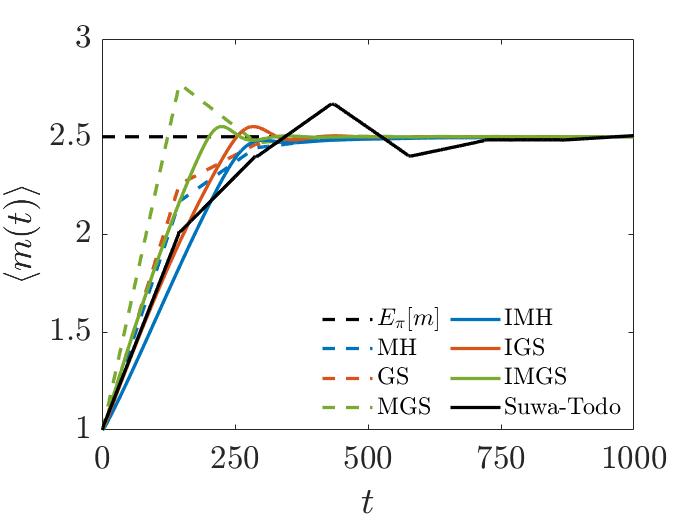}\hfill
\includegraphics[width=.33\textwidth, height=.25\textwidth,]{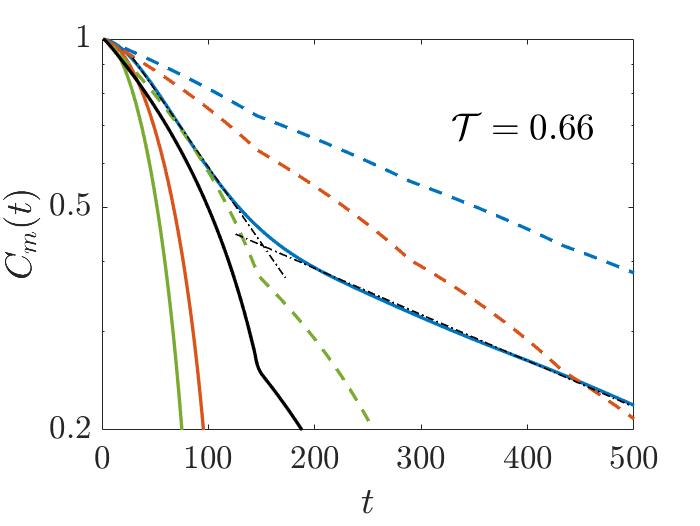}\hfill
\includegraphics[width=.33\textwidth, height=.25\textwidth,]{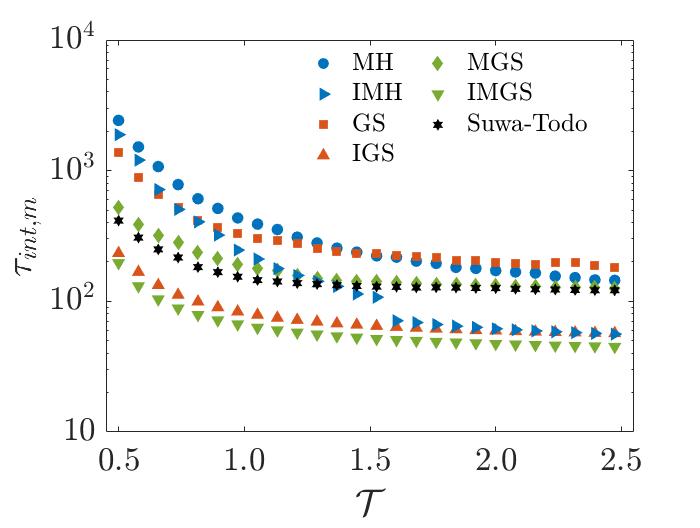}\hfill

\caption{A comparison of IMH, IGS and IMGS algorithms with the deviation parameter set to $\delta = 1$. \textit{\textbf{Left}}: A comparison of the average trajectory of the magnetisation density with respect to time as in eq.\eqref{ensemble average}, $N = 144$ sites, $N_{sim} = 10^5$, $T = 2\times 10^3$ MC-steps, and $\mathcal{T} = 2.0$. \textit{\textbf{Middle}}: The autocorrelation function $C_m(t)$ of the magnetisation density at $\mathcal{T} = 0.66$ obtained from a very long single runs of the algorithms with $T = 10^7$ MC-steps and $N = 144$ sites. The legend is equivalent to the one on the left panel and the black dash-dotted lines project the trajectory of $C_m(t)_{\text{IMH}}$ at the initially fast and then slow decay rate as in eq. \eqref{modes of decay rate}. \textit{\textbf{Right}}: A comparison of $\tau_{int,m}$ at 26 temperatures in the range $\mathcal{T} = 0.5-2.47$ with $N = 144$ sites and $T = 10^7$ MC-steps.}
\label{algorithms comparison}
\end{figure*}

\begin{figure*}[t!]
\centering
\includegraphics[width=.32\textwidth, height=.25\textwidth, ]{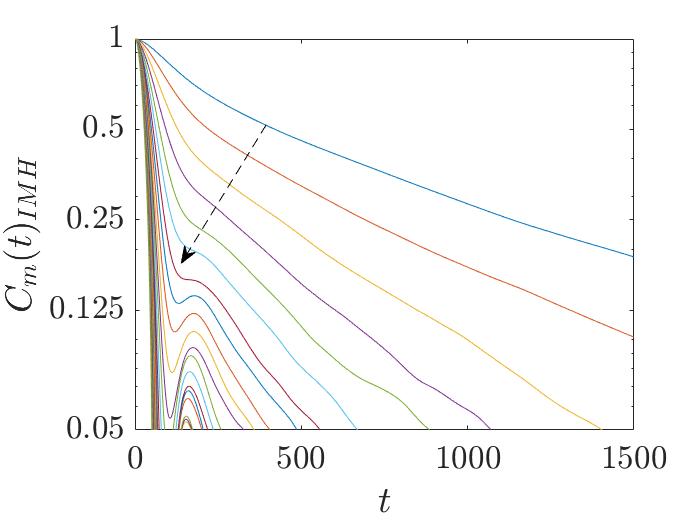}\hfill
\includegraphics[width=.32\textwidth, height=.25\textwidth,]{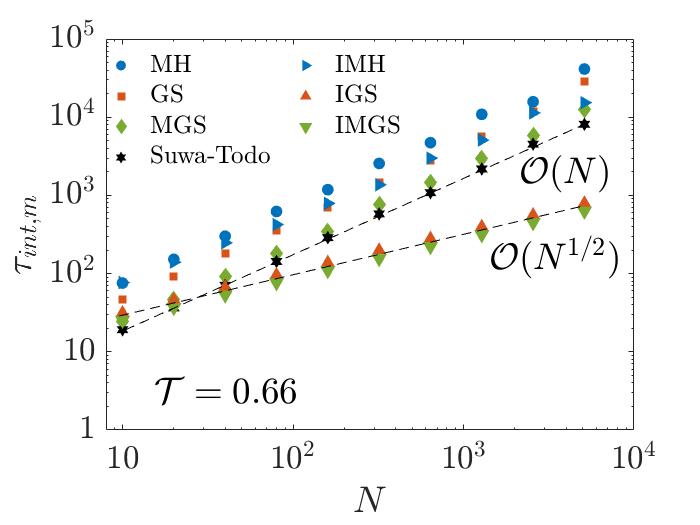}\hfill
\includegraphics[width=.32\textwidth, height=.25\textwidth,]{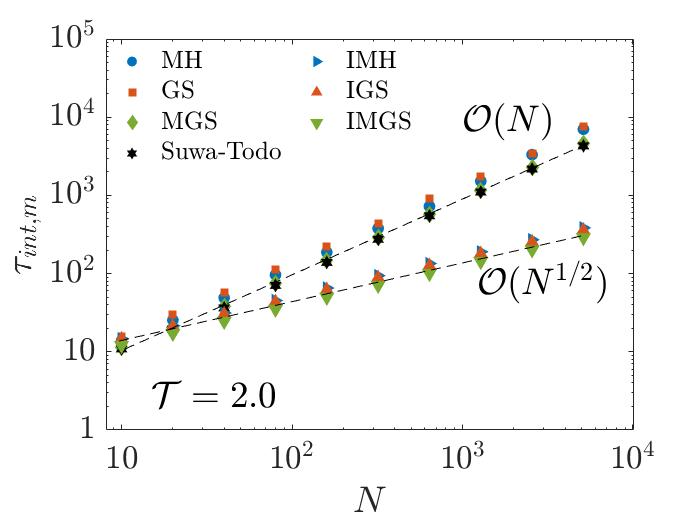}

\caption{\textit{\textbf{Left}}: Trajectories for the autocorrelation functions $C_m(t)_{\text{IMH}}$ for 26 temperatures in the range $\mathcal{T} = 0.5-2.4$, where the direction of the arrow indicates rising temperature. $N = 144$ sites, $T = 10^7$ MC-steps and the deviation parameter is set to $\delta = 1$. Trajectories at higher temperatures are dominated by a fast mode of decay rate $t_{\text{fast}}$ as in eq. \eqref{modes of decay rate}. \textit{\textbf{Middle and right}}: $N$ dependence of $\tau_{int,m}$ at   $\mathcal{T} = 0.66$ and $\mathcal{T} = 2.0$. $\tau_{int,m}$ is obtained from a very long single runs of the algorithms with $T = 10^7$ MC-steps, the dashed lines indicate the best fits. The deviation parameter $\delta$ is set to 1 for IMH, IGS and IMGS.}
\label{IMH autocorrelation times}
\end{figure*}

In Fig.(\ref{algorithms comparison}) and Fig.(\ref{IMH autocorrelation times}) we provide a performance comparison of IMH, IGS and IMGS against each other and some conventional algorithms, namely MH, GS, MGS and the Suwa-Todo algorithm \cite{Suwa-Todo}. In these conventional methods spin sites are updated in sequence, which breaks DBC, and is shown to outperform random updating scheme by reducing autocorrelation times \cite{Ren}. It is clearly seen in the left panel of Fig.(\ref{algorithms comparison}) that no appreciable gain in convergence time is provided by IMH, IGS and IMGS over the conventional methods - except for a gain in convergence time over the Suwa-Todo algorithm. On the other hand  it is evident that IMGS returns the smallest integrated autocorrelation times on the magnetisation density at all given temperatures as shown in the right panel of Fig.(\ref{algorithms comparison}). In particular  we report $\tau_{IMH}/\tau_{IMGS} \sim 6.90$ and $\sim 1.30$ at $\mathcal{T} = 0.66$ and $2.0$ respectively - IMGS seems to outperform IMH by a larger margin at lower temperatures. Such a performance of IMGS is closely followed by the IGS. A particular point of interest is that at all given temperatures both IGS and IMGS return smaller values of $\tau_{int,m}$ than the Suwa-Todo algorithm \cite{Suwa-Todo} - which is considered one of the best local flip algorithms for the Potts model. However $\tau_{int,m}$ for the IMH only becomes shorter than that of the Suwa-Todo algorithm for $\mathcal{T} \geq 1.45$. In particular we report $\tau_{(Suwa-Todo)}/\tau_{IMGS} \sim 2.40$ and $\sim 2.66$ at $\mathcal{T} = 0.66$ and $2.0$ respectively - the integrated autocorrelation times of IMGS are over twice as short compared to those of the Suwa-Todo algorithm.

On the right panel of Fig.(\ref{algorithms comparison}) we also note that at higher temperatures, $\mathcal{T} \geq 1.6$, the IMH algorithm returns $\tau_{int,m}$ values similar to those of IGS, with the two sets of values converging with increasing temperature. At lower temperatures, $\mathcal{T} < 1.6$, while a reduction in $\tau_{int,m}$ is still prevalent for the IMH algorithm, it nonetheless performs relatively poorly as compared to IGS and IMGS. This poor performance at lower temperatures is due to the cross over of $C_m(t)_{\text{IMH}}$ from an initially fast to a slower decay rate as shown in the middle panel of Fig.(\ref{algorithms comparison}). A similar phenomenon is reported for the study of two-dimensional classical XY model with the ECMC algorithm \cite{ECMC continuous spins}. The authors in ref.\cite{ECMC continuous spins} report that the susceptibility autocorrelation function crosses over from an initially fast to a slow decay rate at the criticality. We utilize here a similar description of the autocorrelation function to express $C_m(t)_{\text{IMH}}$ using two time-scales, $t_{\text{fast}}$ and $t_{\text{slow}}$, to characterise the fast and slow modes of decay rates:
\begin{equation}\label{modes of decay rate}
C_m(t)_{\text{IMH}} = A_1\text{exp}\left(-t/t_{\text{fast}}\right) + A_2\text{exp}\left(-t/t_{\text{slow}}\right).
\end{equation}

In the middle panel of Fig.(\ref{algorithms comparison}) we show the autocorrelation functions at $\mathcal{T} = 0.66$. Initially  $C_m(t)_{\text{IMH}}$ decays at a fast time scale $t_{\text{fast}}$ for $t \sim 180$ Monte Carlo steps to $C_m \sim 0.4$, then a cross-over to a slower mode of decay rate $t_{\text{slow}}$ occurs, whereby this new slower decay rate seems characteristic to that of conventional Metropolis-Hastings $C_m(t)_{\text{MH}}$. We observe that increasing the temperature causes the decay rate of $C_m(t)_{IMH}$ to be dominated by the fast time scale $t_{\text{fast}}$ as shown in left panel of Fig.(\ref{IMH autocorrelation times}). A similar cross over between two modes of dacay rate is not observed in IGS and IMGS algorithms - they seem to be well approximated with a single exponential decay. Similar phenomenon whereby a slow diffusive decay succeeds an initial ballistic behaviour has been reported in simulations of particle systems with ECMC algorithms \cite{ECMC Hard disk MC}.
 
The middle and right panel of Fig.(\ref{IMH autocorrelation times}) shows the $N$ dependence of $\tau_{int,m}$ at $\mathcal{T} = 0.66$ and  $\mathcal{T} = 2.0$. For the conventional algorithms $\tau_{int,m}$ scales on the order of $\mathcal{O}(N)$ at both high and low temperatures, whereas in the case of IMH, IGS and IMGS we observe a reduction in the dynamical scaling of $\tau_{int,m}$. At $\mathcal{T} = 2.0$ for all three algorithms, IMH, IGS and IMGS, $\tau_{int,m}$ is on the order of $\mathcal{O}(N^{1/2})$, a square-root reduction of the mixing time. However at the lower temperature of $\mathcal{T} = 0.66$ a different scenario is observed; both IGS and IMGS still provide a square root reduction of the mixing time, and IMH now only scales on the order of $ \sim \mathcal{O}(N^{0.85})$. The square root reduction of the mixing time was shown to be optimal through the lifting framework \cite{Chen}, and it therefore seems that at sufficiently high temperatures all three algorithms, IMH, IGS and IMGS present a maximal improvement of mixing time. However, at a low temperature, only IGS and IMGS retain the best mixing time achievable. 
 
\subsubsection*{Energy density as the lifting coordinate}

\begin{figure*}[t!]
\centering

\includegraphics[width=.32\textwidth, height=.25\textwidth, ]{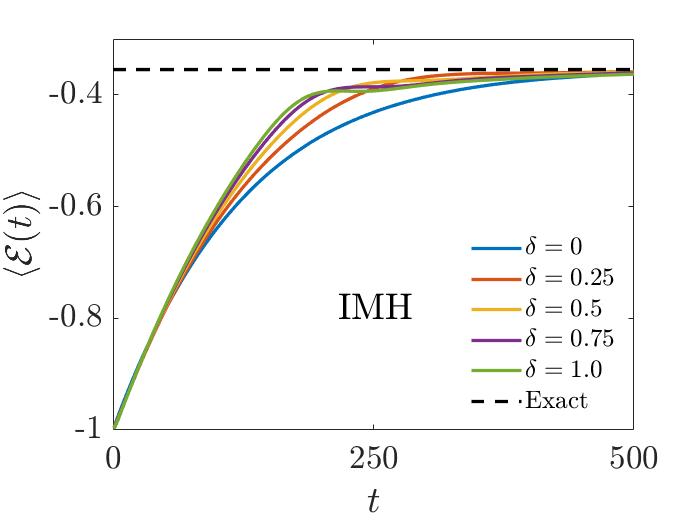}\hfill
\includegraphics[width=.32\textwidth, height=.25\textwidth, ]{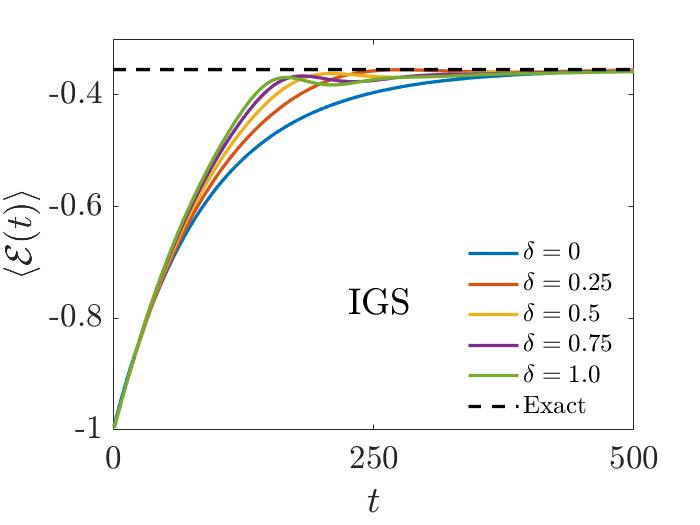}\hfill
\includegraphics[width=.32\textwidth, height=.25\textwidth, ]{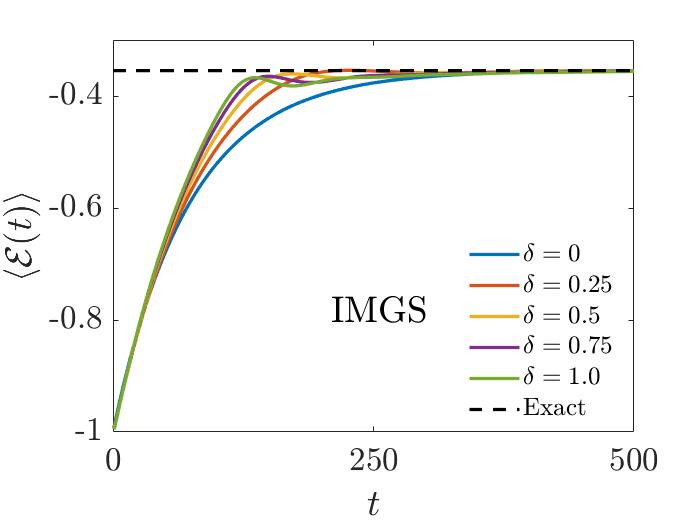}

\caption{The average trajectories tracing the evolution of energy density $\mathcal{E}$ with respect to time starting from the initial conditions, the averages have been computed using eq.\eqref{ensemble average} and the exact value for the equilibrium energy density, $\mathcal{E} \simeq -0.3547$, using eq. \eqref{exact energy density}. The simulations were initialised with $\sigma_k = 1$ for $k = 1,...,N$ and a random assignment of $\varepsilon \in \lbrace +1,-1 \rbrace$. $N = 144$, $\mathcal{T} = 2.0$, $N_{\text{sim}} = 10^5$, and $T = 2 \times 10^3$. The deviation parameter $\delta$ indicates deviation from DBC.}
\label{energy convergence}
\end{figure*}

In this section we take the lifting coordinate $f$ in the skewness function in \eqref{skewness adapted}, to be the energy density $\mathcal{E}$ of the 1D 4-state Potts model. Imposing periodic boundary conditions and setting $J_{k,k+1} = J$ for $k = 1,...,N$ allows us to write the energy density of the system in the form:
 \begin{equation}
 \mathcal{E} = -\frac{J}{N}\sum\limits_{k = 1}^{N} \delta\left(\sigma_k, \sigma_{k+1} \right),
 \end{equation}
where $\delta(\cdot)$ here denotes the Kronecker delta function, not to be confused with the parameter in the skewness function given in \eqref{skewness adapted}. In Figure.(\ref{energy convergence}) we show the average trajectories tracing the evolution of energy density with respect to time at $\mathcal{T} = 2.0$. The exact value for the equilibrium energy density of the model can be analytically deduced from its partition function and is given by
 \begin{equation}\label{exact energy density}
 \mathcal{E} = -J\frac{e^{\beta J}}{e^{\beta J} - 1 + q}.
 \end{equation} 
In Fig.(\ref{energy convergence}) all trajectories converge on the exact value, but deviation from the DBC seems to induce an initially fast convergence rate in all three algorithms.

\begin{figure*}[t!]
\centering

\includegraphics[width=.32\textwidth, height=.25\textwidth, ]{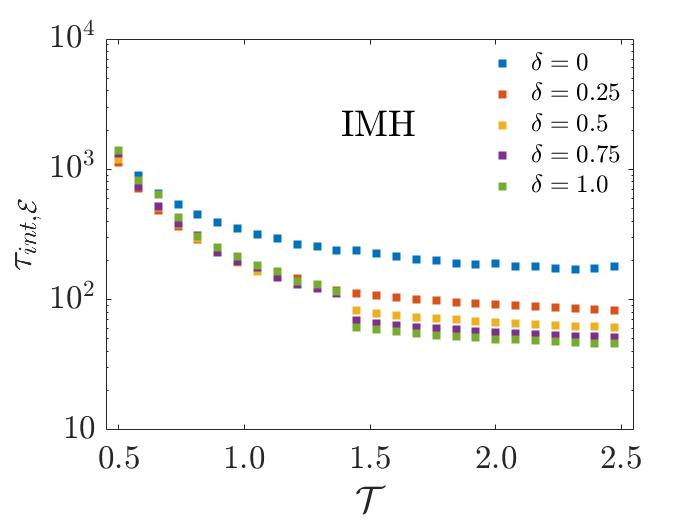}\hfill
\includegraphics[width=.32\textwidth, height=.25\textwidth, ]{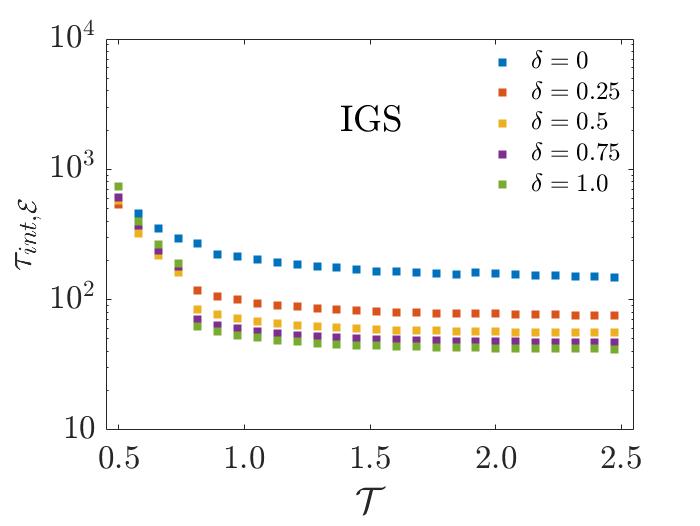}\hfill
\includegraphics[width=.32\textwidth, height=.25\textwidth, ]{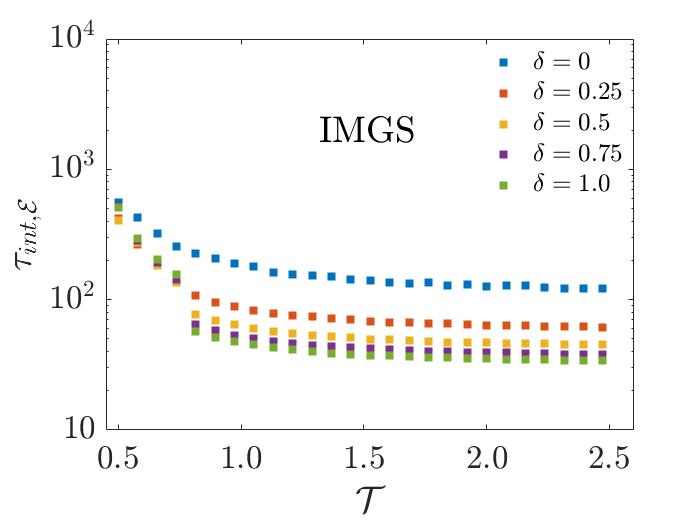}

\caption{Integrated autocorrelation times $\tau_{int, \mathcal{E}}$ for energy density $\mathcal{E}$ at 26 temperatures in the range $\mathcal{T} = 0.5-2.47$. The deviation parameter $\delta$ indicates deviation from DBC.  The values are obtained from a very long single runs of the algorithms with $T = 10^7$ MC-steps using $N = 144$ sites. The deviation parameter $\delta$ indicates deviation from DBC.}
\label{IAC vs Temp for energy density}
\end{figure*}

The integrated autocorrelation time for energy density is computed for 26 temperatures in the range $\mathcal{T} = 0.5-2.47$, we show this in Fig.(\ref{IAC vs Temp for energy density}). The pattern observed is very similar to that in Fig.(\ref{tau_vs_kT_magnetisation}): deviation from the DBC induces reduction in $\tau_{int,\mathcal{E}}$ in all three algorithms.

\begin{figure*}[t!]
\centering

\includegraphics[width=.32\textwidth, height=.25\textwidth, ]{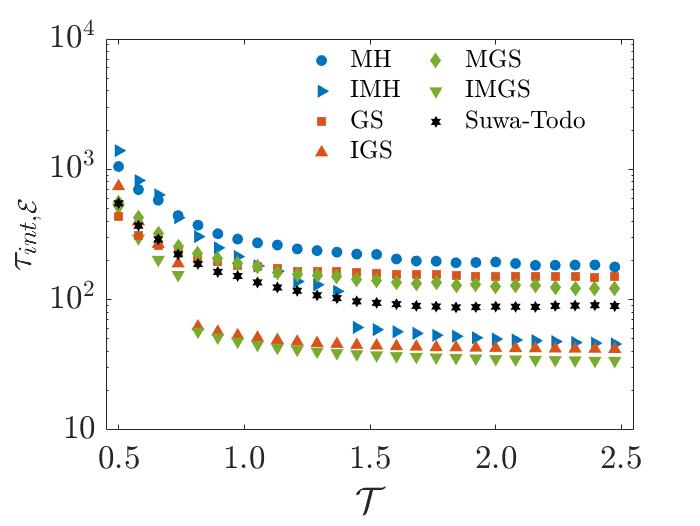}\hfill
\includegraphics[width=.32\textwidth, height=.25\textwidth, ]{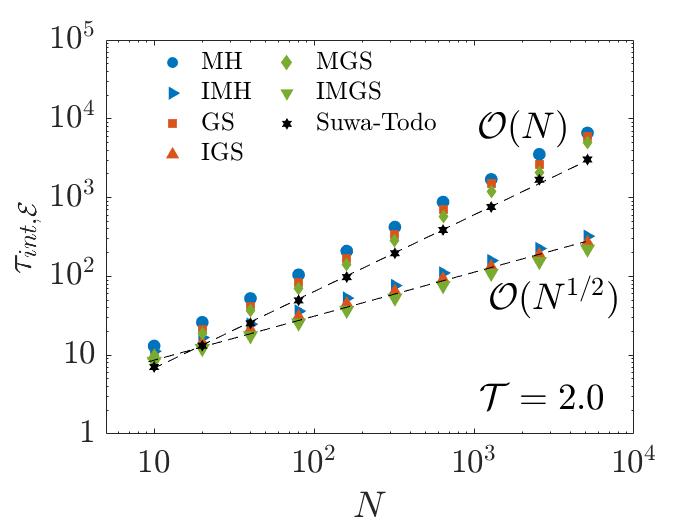}\hfill
\includegraphics[width=.32\textwidth, height=.25\textwidth, ]{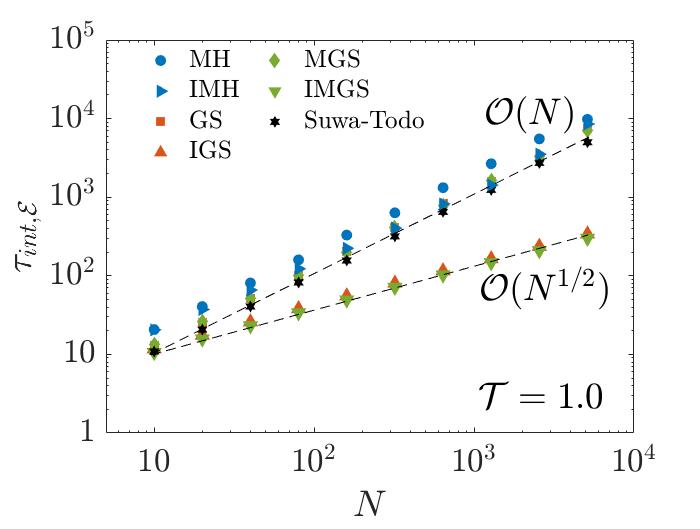}
\caption{\textit{\textbf{Left}}: A comparison of $\tau_{int,\mathcal{E}}$ at 26 temperatures in the range $\mathcal{T} = 0.5-2.47$, $\tau_{int,\mathcal{E}}$ is obtained from a very long single runs of the algorithms with $T = 10^7$ MC-steps and $N = 144$ sites. For IMH, IGS and IMGS the deviation parameter $\delta$ is set to 1. \textit{\textbf{Middle and Right}}: $N$ dependence of $\tau_{int,\mathcal{E}}$ at $\mathcal{T} = 2.0$ and $1.0$ respectively. $\tau_{int,\mathcal{E}}$  is obtained from a very long single runs of the algorithms with $T = 10^7$ MC-steps. For IMH, IGS and IMGS the deviation parameter $\delta$ is set to 1.}
\label{energy scaling comparison}
\end{figure*}

\noindent However we point out that for both IGS and IMGS, at temperatures of $\mathcal{T} < 0.8$, the autocorrelation functions for energy density exhibit decay rates at two time scales, $t_{\text{fast}}$ and $t_{\text{slow}}$ as dictated in \eqref{modes of decay rate}. This is not the case with their respective autocorrelation functions for magnetisation density, which seem to be well described by a single decay rate. Unlike $\tau_{int,m}$ the reduction in $\tau_{int,\mathcal{E}}$ therefore shows a drastic degradation at low temperatures for both IGS and IMGS.

A comparison of $\tau_{int,\mathcal{E}}$ to those obtained from conventional algorithms is shown on the left panel of Fig.(\ref{energy scaling comparison}). It seems that, concerning $\tau_{int,\mathcal{E}}$, the optimum superiority of both IGS and IMGS over IMH is in the temperature window of $0.7 < \mathcal{T} < 1.4$.   The right panels of Fig.(\ref{energy scaling comparison}) therefore show that at a high temperature $\tau_{int,\mathcal{E}}$ scales on the order of $\mathcal{O}(N^{1/2})$ for IMH, IGS and IMGS, but this square root reduction of the mixing time is only retained by IGS and IMGS at low temperatures.

These results show that by setting the lifting coordinate $f$ in the skewness function in \eqref{skewness adapted} as the observable of interest, IMH, IGS and IMGS can significantly reduce the integrated autocorrelation times of this particular observable in comparison to conventional algorithms. The IMGS in particular provides the best performance of the three methods.

\subsection*{Extensions to continuous state systems}

As a simple example of possible applications to continuous state systems, one may consider using Monte Carlo simulations to construct a free energy profile of a system described by a symmetrical 1D double well potential  
\begin{equation}\label{model potential equation}
U(x) = C(x -1)^2(x+1)^2,
\end{equation}
where $C \geq 0$ is a tunable constant. The potential has two minima at $x = -1$ and $x = 1$ and an energy barrier of magnitude $C$ at $x = 0$. Let us consider a Metropolis-Hastings Monte Carlo scheme using a Gaussian proposal in the $x$-coordinate given by
\begin{equation}\label{1D potential propose cont}
x_{i+1}' = x_i + \varsigma\xi,
\end{equation}
where $\xi = \mathcal{N}(0,1)$ and $\varsigma$ is the standard deviation set to 0.11. $x'$ indicates the proposal for the next Monte Carlo time step $(i+1)$. We will henceforth refer to this particular Metropolis-Hastings scheme with the proposal given in \eqref{1D potential propose cont} as MH-C to indicate the \emph{continuous} states in the $x$-coordinate.

Alternatively one may consider discrete points along the $x-$coordinate linearly spaced with a space width of $\Delta$ and wish to propose the next state in the Monte Carlo time step according to
\begin{equation}\label{1D potential propose}
x_{i+1}' = x_i + \mathcal{U}\lbrace -n,-n+1,...,-1,1, ... ,n-1,n \rbrace \Delta,
\end{equation} 
where $\mathcal{U}$ is the discrete uniform distribution and $n$ determines the maximum deviation from the current state $x_i$. Setting the spacing $\Delta$ as infinitesimally small a discrete state approximation of a continuous state space can be realised along the $x-$coordinate. In what follows we will refer to the standard Metropolis-Hastings algorithm with the proposal in \eqref{1D potential propose} as simply MH. 

Finally we point out that the irreversible counterpart of MH with SDBC can now be constructed by simply following the general recipe in Algorithm.\ref{IMH algo}. The proposal in \eqref{1D potential propose}, $(x',\varepsilon \vert x, \varepsilon)$, is accepted with the probability: $ \Theta(x,x',\varepsilon)A(x' \vert x)_{\text{MH}}$, where upon rejection one accepts the state $(x,-\varepsilon \vert x, \varepsilon)$ with probability: $\Lambda(x,-\varepsilon \vert x, \varepsilon)/(1 - \sum_{x' \neq x}T(x', \varepsilon \vert x, \varepsilon))$, or else the current state $(x,\varepsilon)$ is retained. In the transition probability $T(x', \varepsilon \vert x, \varepsilon) = \Theta(x,x',\varepsilon)T(x' \vert x)_{\text{MH}}$, the MH transition $T(x' \vert x)_{\text{MH}}$ and acceptance $A(x' \vert x)_{\text{MH}}$ are defined in equations \eqref{metropolis transition matrix1} and \eqref{metropolis acceptance prob} respectively. Two points merit attention here; firstly, the evaluation of the probability to switch replica $\varepsilon \rightarrow -\varepsilon$ requires summation over $2n$ states, in practice this summation is continually updated and for very large $n$ the computation can be easily done in parallel. Secondly, the skewness function $\Theta(x,x',\varepsilon)$, while it must satisfy both conditions \eqref{skewness condition 1} and \eqref{skewness condition 2}, should ideally utilize the physics of the system in question, so for example for the system under our consideration with the potential energy defined in \eqref{model potential equation}, for efficient sampling we want to enhance crossing of the energy barrier between the two minima. As a convenient choice we make use of the skewness function in \eqref{skewness adapted} and set the lifting coordinate $f$ to be the state of the system along the $x-$coordinate, so that 
\begin{equation}\label{1D model skewness}
\Theta(x,x',\varepsilon) =  \varphi\left(1 + \delta\varepsilon \, \text{sgn}(x' - x)\right).
\end{equation} 
In this manner the acceptance probability is biased such that in the $\varepsilon = +1$ replica Monte Carlo moves that drive the system towards increasing $x-$coordinate are more likely to be accepted than those that propose to decrease the $x-$coordinate, while the opposite is true in $\varepsilon = -1$ replica. In a given replica the system therefore acquires momentum in a specific direction in the $x$-coordinate, thus barrier crossing between the energy minima is expected to be more rapid.

\begin{figure*}[t!]
\centering

\includegraphics[width=.32\textwidth, height=.25\textwidth, ]{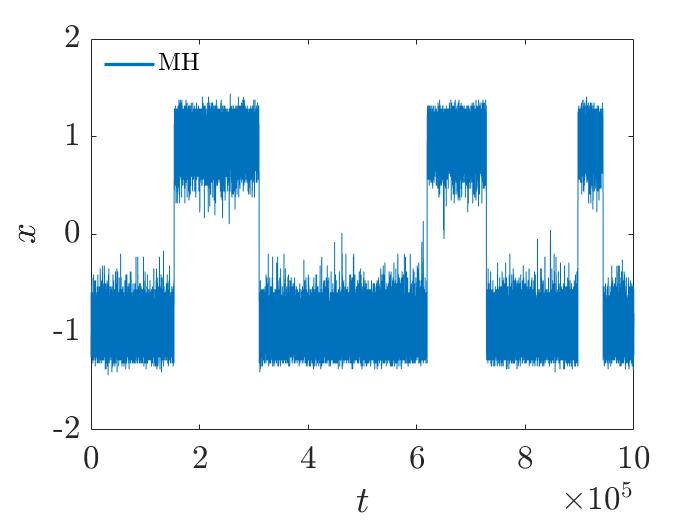}\hfill
\includegraphics[width=.32\textwidth, height=.25\textwidth, ]{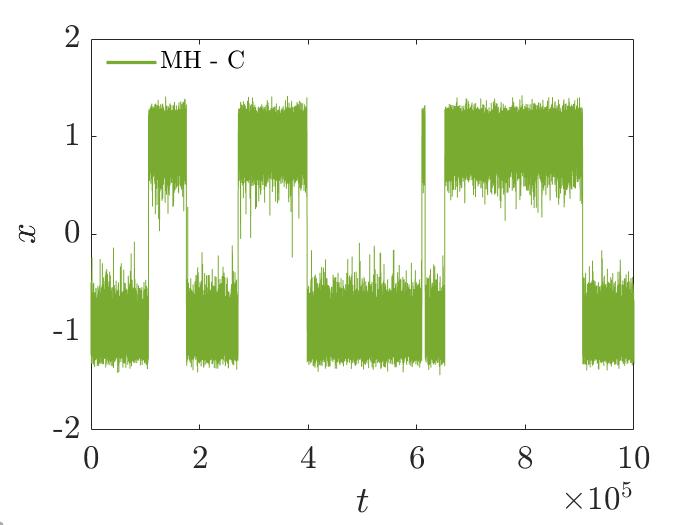}\hfill
\includegraphics[width=.32\textwidth, height=.25\textwidth, ]{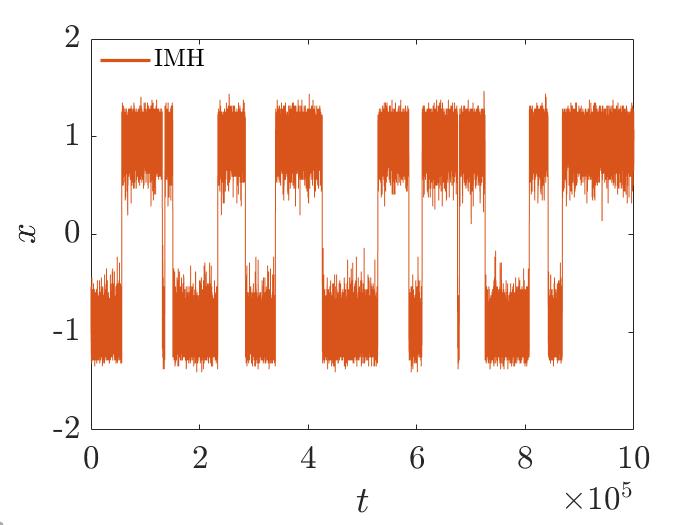}

\includegraphics[width=.32\textwidth, height=.25\textwidth, ]{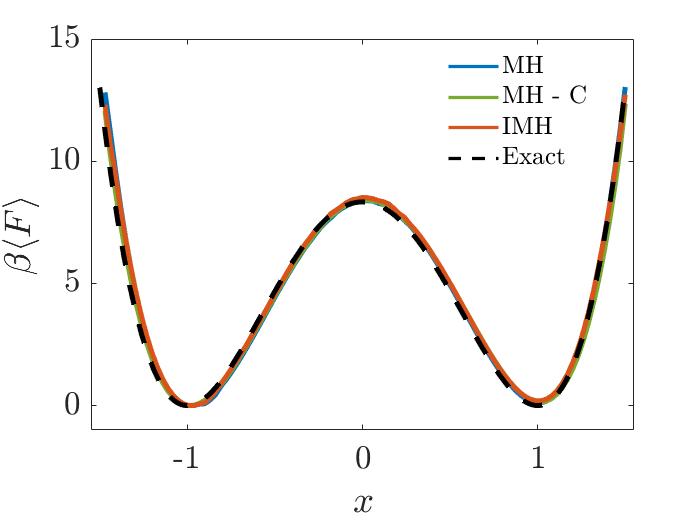}\hfill
\includegraphics[width=.32\textwidth, height=.25\textwidth, ]{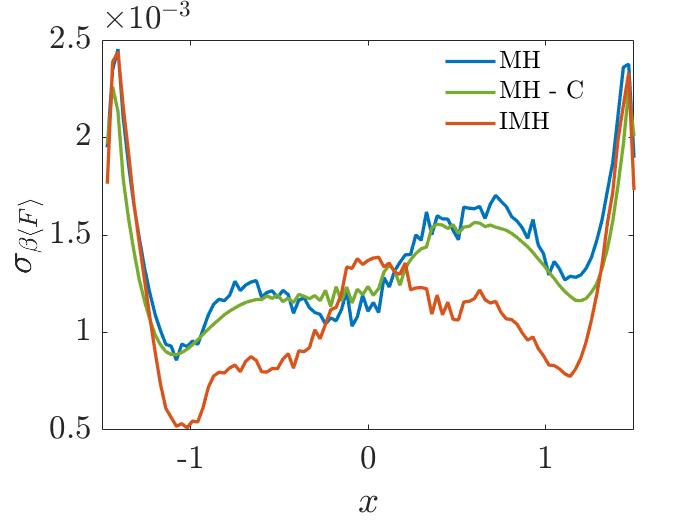}\hfill
\includegraphics[width=.32\textwidth, height=.25\textwidth, ]{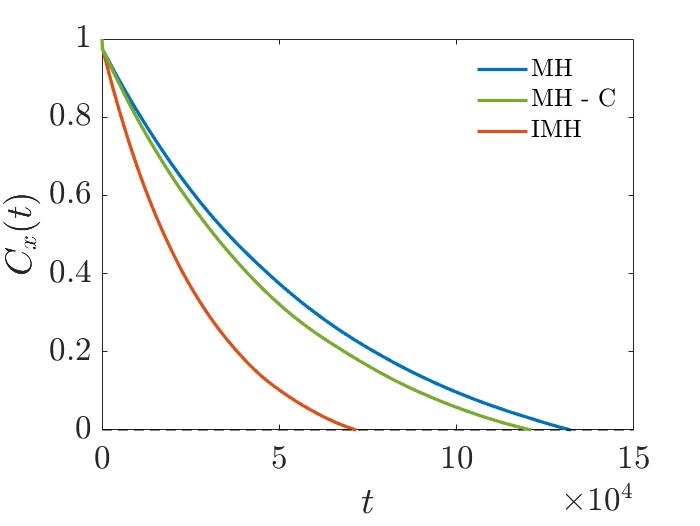}

\caption{Simulation of a simple system with 1D model potential as given in \eqref{model potential equation} with MH, MH-C and IMH algorithms and simulation parameters: $\beta \sim 0.83$, $T = 10^7$ MC-steps and $N_{sim} = 450$ independent simulations. \textbf{Top row}: Time evolution of the $x-$coordinate obtained from single runs of the algorithms. \textbf{Bottom left}: The average trajectory of free energy profile $\beta F(x) = -\text{ln}(\pi(x))$ as in eq. \eqref{ensemble average}. \textbf{Bottom center:} Standard error on the average trajectory of free energy profile. \textbf{Bottom right:} Autocorrelation function $C_x(t)$ of observable $x$.}
\label{1D model potential}
\end{figure*}

We show some results of the simulations with MH, MH-C and IMH in Fig.(\ref{1D model potential}). For both MH and IMH we had set $n = 5$ and the space width $\Delta = 0.0303$ in \eqref{1D potential propose} so that the average step-size (i.e. deviation from the current state) in the proposal corresponds roughly to that of MH-C for fair comparison. All three algorithms converge to the correct target distribution $\pi(x)$ as is clearly evident from the free energy profile $\beta F(x) = -\text{ln}(\pi(x))$ shown on the bottom left of Fig.(\ref{1D model potential}). The top row of Fig.(\ref{1D model potential}) shows the evolution of the $x-$coordinate with respect to time for a single run of the algorithms, we observe the expectedly similar performance of MH and MH-C, whereas IMH clearly exhibits a ballistic behaviour with more rapid crossing of the energy barrier. The superior mobility of IMH along the reaction coordinate can induce faster convergence to the stationary distribution in comparison to its reversible counterparts, MH and MH-C. This seems a typical advantage of the lifting framework as breaking DBC can accelerate the otherwise diffusive exploration of the reaction coordinate \cite{Diaconis,Turitsyn,hot topic 7}. We observe on the bottom center of Fig.(\ref{1D model potential}) that IMH returns the smallest standard errors on the free energy profile on either side of the energy barrier. The autocorrelation functions for the observable $x$ (bottom right of Fig.(\ref{1D model potential})) indicate that $\tau_{int,x}$ is $\sim 1.8$ times shorter for IMH compared to MH-C, the asymptotic variance on the lifting coordinate can therefore be appreciably reduced with the IMH algorithm.

We have provided here a simple example to demonstrate that a  skewness function can be carefully constructed to optimize sampling of the state space with the IMH algorithm. The increased mobility in state space can be attributed with the optimal choice of the skewness function, e.g. for the 1D model potential discussed here our observable of interest is states along the $x-$coordinate; as effective sampling of this state space is of interest to construct the corresponding free energy profile. Therefore in the interest of increasing mobility along the $x-$coordinate we had provided in \eqref{1D model skewness} a skewness function that sets the state $x$ as the lifting coordinate. We cautiously state that an optimal skewness function should in general utilize the observable of interest as the lifting coordinate.

We also point out that the simple 1D example here can be extended to 3D cases that may be dictated by more irregular potential energy profiles. Furthermore, while both the IGS and IMGS algorithms are applicable to general systems with discrete state space, their application to continuous state systems, as with the IMH algorithm, can be practically feasible with a discrete state space approximation. We have demonstrated with the simple example of a 1D model potential that the IMH algorithm can be successfully adapted for a Monte Carlo simulation of a continuous state system with improved performance than the standard Metropolis-Hastings with DBC. With similar reasoning the application of IGS and IMGS can be practically viable, e.g. the IGS transition rate in \eqref{IGS} may be adapted to 
\begin{align}
\mathcal{G}(x', \varepsilon \vert x, \varepsilon) &= \Theta(x,x',\varepsilon)G(x' \vert x)\,\,\,\,\,\ \forall \,\,\, x' \neq x, \nonumber  \\
\mathcal{G}(x, \varepsilon \vert x, \varepsilon) &= 1 - \sum\limits_{x' \neq x}\mathcal{G}(x', \varepsilon \vert x, \varepsilon), \nonumber 
\end{align}
where the Gibbs transition
\begin{equation}
G(x' \vert x) = \frac{\pi(x')}{\sum\limits_{m = -n}^{n}\pi(x + m\Delta)}\,\,\,  \forall \,  x' \in \lbrace x-n\Delta,..., x-\Delta, x , x+\Delta,..., x + n\Delta \rbrace \nonumber
\end{equation}
to state $x'$ could be limited to $2n+1$ states, i.e. $2n$ states in the vicinity of the current state $x$, as for example in equation \eqref{1D potential propose}. The irreversible algorithms presented here may therefore be useful in constructing free energy landscapes of more complex systems with higher dimensionality, as for example bio-molecular systems.

\section*{Discussion} 

In summary, we have presented in this paper three algorithms on the basis of SDBC, namely the irreversible Metropolis-Hastings (IMH), irreversible Gibbs sampler (IGS) and irreversible Metropolized-Gibbs sampler (IMGS). The IMH presented here is a generalisation, to classical spin systems, of the prototype algorithm presented by Turitsyn et.al. for the mean field Ising model \cite{Turitsyn}, our generalisation now makes it applicable to classical spin systems in general. We have managed this generalisation by building on the works of Sakai and Hukushima on the 2D Ising model \cite{Sakai Hukushima 2D}, specifically by the adaptation of the skewness function $\Theta^{(\varepsilon)}_{ij}$, which characterises the violation of DBC, so that it may now use any generic observable $f$ as the lifting coordinate. Performance analysis of IMH on 1D 4-state Potts model indicate a square-root reduction of the mixing time at high temperatures, while performance at low temperatures remains modest.

The IGS and IMGS presented in this paper are respectively the irreversible counterparts with SDBC of the random-scan Gibbs sampler \cite{Gibbs sampler} and the random-scan Metropolized-Gibbs sampler \cite{Liu2}. We have presented these two algorithms in general formulation so as to be applicable to any system with discrete degrees of freedom. Performance analysis on 1D 4-state Potts model show that both IGS and IMGS return shorter autocorrelation times in comparison to IMH and some conventional algorithms. The integrated autocorrelation times for magnetisation and energy density scale on the order of $\mathcal{O}(N^{1/2})$ at both high and low temperatures, as compared to conventional algorithms which scale on the order of $\mathcal{O}(N)$. This square-root reduction of the mixing time may be the optimal improvement achievable through the lifting framework \cite{Chen}. 

\begin{figure*}[t!]
\centering

\includegraphics[width=.45\textwidth, height=.35\textwidth, ]{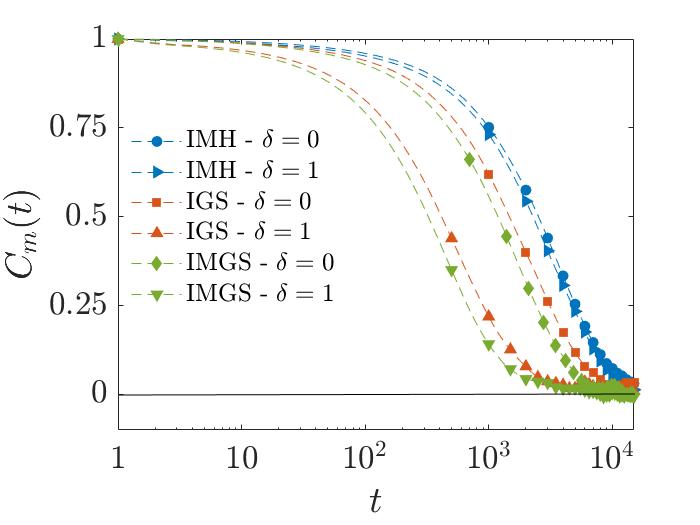}\hfill
\includegraphics[width=.45\textwidth, height=.35\textwidth, ]{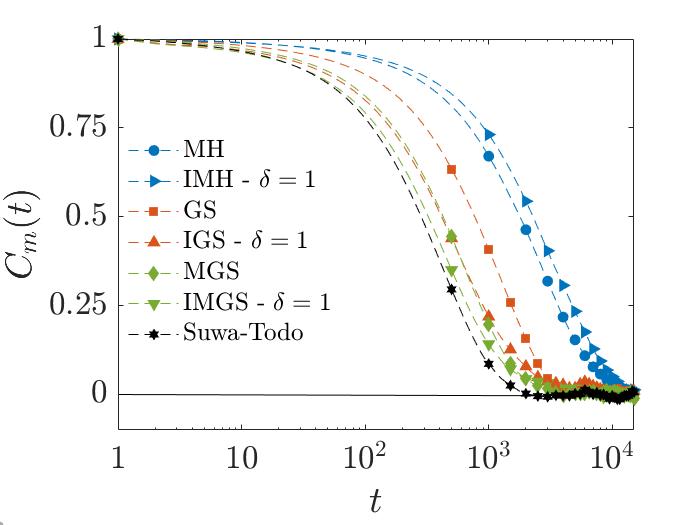}

\caption{Autocorrelation functions of magnetisation density $C_m(t)$ for the 2D 4-state Potts model at the critical temperature $\mathcal{T} \simeq 0.910$. The autocorrelation functions are obtained from a very long single runs of the algorithms with $T = 10^7$ sweeps. The system size is $16\times 16$. \textbf{Left:} DBC is recovered by setting the deviation parameter $\delta$ to 0 while $\delta = 1$ characterises maximum deviation from DBC. \textbf{Right:} We show comparison of $C_m(t)$ with conventional algorithms namely MH, GS, MGS and the Suwa-Todo algorithm whereby in these conventional algorithms spins are updated in sequence.}
\label{2D Potts figure}
\end{figure*}

To further test the efficiency of IMH, IGS and IMGS, large scale simulations of 2D and 3D discrete state spin systems at the criticality is of interest. Preliminary simulation results on a 2D 4-state Potts model of size $16^2$ at the criticality indicate that both IGS and IMGS provide shorter autocorrelation times in comparison to their respective reversible counterparts that satisfy the strict DBC, this is shown in the left panel of Fig.(\ref{2D Potts figure}); notice that the IMH does not perform significantly better than its reversible counterpart. The autocorrelations of the magnetisation density are also compared to those from conventional algorithms namely, MH, GS, MGS and the Suwa-Todo algorithm \cite{Suwa-Todo}, where in these conventional methods spins are updated in sequence, which breaks DBC. As clearly seen in the right panel of Fig.(\ref{2D Potts figure}) the IMGS seems second in performance only to the Suwa-Todo algorithm. In particular the integrated autocorrelation time of IMGS is 4.8 times shorter than that by sequential-scan MH, 1.9 times than the sequential-scan Gibbs sampler and a modest 1.1 times shorter than the sequential-scan MGS. In their current form IMH, IGS and IMGS ensure invariance of the target distribution only with random updating scheme. Sequential updating schemes have however been shown to reduce autocorrelation times \cite{Ren}. Our current work in progress \cite{Fahim and George} therefore looks at implementing IMGS with sequential updating scheme.      

Both IGS and IMGS are applicable to general systems with discrete degrees of freedom, it is therefore of interest, for a future study, to test the performance of these algorithms in the study of more complicated statistical-physics models, such as the Potts spin glass models.  In addition, the lifting framework with SDBC can be applied to generalized-ensemble algorithms in an attempt to improve their efficiency, for example, recent application of the lifting technique was applied  to the updating scheme of inverse temperature in simulated tempering \cite{Sakai Hukushima simulated tempering} with improved efficiency over the standard updating scheme with DBC. In our current work in progress \cite{Fahim and George 2} we are implementing the IGS and IMGS in the updating scheme of inverse temperature in simulated and parallel tempering.

Spin-cluster algorithms\cite{Swendsen and Wang,Wolff} offer alternative, very efficient Monte Carlo sampling of simple spin systems, which have been superior to the study of critical phenomena compared to the conventional (Metropolis-type) Monte Carlo methods - particularly in the suppression of critical slowing down \cite{MCMC physics2}. However, broader applications of cluster algorithms to off-lattice systems remain elusive. Only the geometric cluster algorithms (GCA) \cite{GCA1,GCA2} were recently developed for fluid models that may offer efficient general alternatives. Classical spin-cluster algorithms, while very impressive, remain confined to a few spin models. For example, the accelerated dynamics brought about by cluster algorithms in ferromagnetic spin models remain difficult to replicate in the more disordered generic spin glass models \cite{spin glass fail}.  Therefore, here we have focused on the development of more broadly applicable conventional Monte Carlo methods, albeit combination with GCA-type algorithms and other cluster approaches remains of interest. For example, the spin-cluster algorithms are reported to be ineffective in the simulation of 3D XY spin glass models \cite{ECMC continuous spins}, while the ECMC algorithm (which combines the concept of lifting with the factorized Metropolis filter) can outperform both the conventional Metropolis-Hastings algorithm and spin-cluster methods \cite{ECMC continuous spins}. Future work could therefore explore the possibility of combining IMH, IGS and IMGS with cluster algorithms.

In this paper we have made use of the \textit{inter-replica} transition probability $\Lambda_i^{(\varepsilon)}$ of the TCV type \cite{Turitsyn} as described in \eqref{TCV type}, we remark that the choice of $\Lambda_i^{(\varepsilon)}$ is not restricted but several other choices have been proposed and studied analytically and numerically \cite{Sakai Hukushima 1D, Sakai Hukushima eigenvalue}. The efficiency of the algorithms here are dictated by the choice of $\Lambda_i^{(\varepsilon)}$, it is therefore of interest to consider the behaviour of IGS and IMGS with alternative choice of \textit{inter-replica} transition probability. 

Furthermore, we have discussed that the skewness function, $\Theta^{(\varepsilon)}_{ij}$, requires a careful construction so as to impose the SDBC condition on the extended transition matrix, however within the bounds of requirement \eqref{skewness condition 1}  and \eqref{skewness condition 2}, it is possible to engineer an appropriate skewness function that may provide an optimum efficiency of the algorithm for a given system. The skewness function presented here, eq.\eqref{skewness adapted}, readily utilizes a generic observable $f$ as the lifting coordinate and is therefore broadly applicable. A careful selection of the lifting coordinate $f$ may therefore provide a more effective sampling of the state space in the Monte Carlo study of bio-molecular systems, such as proteins, which are prone to being stuck in local minimum energy states.

\begin{acknowledgement}

F.F. is supported by the EPSRC Centre for Doctoral Training in Cross-Disciplinary Approaches to Non-Equilibrium Systems (EPSRC reference: EP/L015854/1). E.R. acknowledges support from EPSRC (EP/R013012/1) and the ERC (Project No. 757850 BioNet).

\end{acknowledgement}


\begin{thebibliography}{9}
\bibitem{MCMC physics1} Landau, D. P.; Binder, K. \textit{A Guide to Monte Carlo Simulations in Statistical Physics}, 2nd ed.; Cambridge University Press: Cambridge, 2005. 

\bibitem{MCMC physics2} Newman, M. E. J.; Barkema, G.T. \textit{ Monte Carlo methods in Statistical Physics}; Oxford University Press: New York, 2001.

\bibitem{MCMC biochemistry} Hansmann, U. H. E.; Okamoto, Y. New Monte Carlo algorithms for protein folding. \textit{Curr. Opin. Struct. Biol.} \textbf{1999}, \textit{9}, 177-183.

\bibitem{MCMC biochemistry2} Kolinski, A.; Skolnick, J. Monte carlo simulations of protein folding. II. Application to protein A, ROP, and crambin. \textit{Proteins: Struct, Funct, Bioinf.} \textbf{1994}, \textit{18},  353-366.

\bibitem{MCMC finance} Shonkwiler, R. W. \textit{Finance with Monte Carlo}; Springer: New York, 2013. 

\bibitem{Metropolis} Metropolis, N.; Rosenbluth, A. W.; Rosenbluth, M. N.; Teller, A. H.; Teller, E. Equation of State Calculations by Fast Computing Machines. \textit{J. Chem. Phys.} \textbf{1953}, \textit{21}, 1087. 

\bibitem{Gibbs sampler} Geman, S.; Geman, D. Stochastic Relaxation, Gibbs Distributions, and the Bayesian Restoration of Images. \textit{IEEE Trans. Pattern Anal. Mach. Intell.} \textbf{1984}, \textit{6}, 721-741. 

\bibitem{Barker} Barker, A. A. Monte Carlo calculations of the radial distribution functions for a proton-electron plasma.  \textit{Aust. J. Phys.} \textbf{1965}, \textit{18}, 119-134. 

\bibitem{Generalized ensemble} Mitsutake, A.; Sugita, Y.; Okamoto, Y. Generalized-ensemble algorithms for molecular simulations of biopolymers. \textit{Biopolymers} \textbf{2001}, \textit{60}, 96-123.

\bibitem{REM 1} Swendsen, R. H.; Wang, J. S. Replica Monte Carlo Simulation of Spin-Glasses. \textit{Phys. Rev. Lett.} \textbf{1986}, \textit{57}, 2607.

\bibitem{REM 3} Hansmann, U. H. E. Parallel tempering algorithm for conformational studies of biological molecules. \textit{Chem. Phys. Lett.} \textbf{1997}, \textit{281}, 140-150.

\bibitem{REM 4} Falcioni, M.; Deem, M. W. A biased Monte Carlo scheme for zeolite structure solution. \textit{J. Chem. Phys.} \textbf{1999}, \textit{110}, 1754.

\bibitem{REM 5} Sugita, Y.; Okamoto, Y. Replica-exchange molecular dynamics method for protein folding. \textit{Chem. Phys. Lett.} \textbf{1999}, \textit{314}, 141-151.

\bibitem{STM} Marinari, E.; Parisi, G. Simulated Tempering: A New Monte Carlo Scheme. \textit{EPL} \textbf{1992}, \textit{19}, 451-458.

\bibitem{MUCA 1} Berg, B. A.; Neuhaus, T. Multicanonical algorithms for first order phase transitions. \textit{Phys. Lett. B.} \textbf{1991}, \textit{267}, 249-253.

\bibitem{MUCA 2} Berg, B. A.; Neuhaus, T. Multicanonical ensemble: A new approach to simulate first-order phase transitions. \textit{Phys. Rev. Lett.} \textbf{1992}, \textit{68}, 9-12.

\bibitem{Berg} Berg, B. A. Introduction to Markov Chain Monte Carlo Simulations and their Statistical Analysis. 2004, arXiv:cond-mat/0410490. arXiv e-prints. https://arxiv.org/abs/cond-mat/0410490 (accessed: Jan 27, 2020). 

\bibitem{Swendsen and Wang} Swendsen, R. H.; Wang, J. S. Nonuniversal critical dynamics in Monte Carlo simulations.  \textit{Phys. Rev. Lett.} \textbf{1987}, \textit{58}, 86. 

\bibitem{Wolff} Wolff, U. Collective Monte Carlo Updating for Spin Systems. \textit{Phys. Rev. Lett.} \textbf{1989}, \textit{62}, 361. 

\bibitem{Peskun} Peskun, P. H. Optimum Monte-Carlo sampling using Markov chains. \textit{Biometrika} \textbf{1973}, \textit{60}, 607-612.

\bibitem{Liu2} Liu,  J. S. Peskun's theorem and a modified discrete-state Gibbs sampler. \textit{Biometrika} \textbf{1996}, \textit{83}, 681-682.
 
\bibitem{Liu1} Liu, J. S. Metropolized independent sampling with comparisons to rejection sampling and importance sampling. \textit{Stat Comput} \textbf{1996}, \textit{6}, 113-119.

\bibitem{Metropolized-Gibbs} Pollet, L.; Rombouts, S. M. A.; Van Houcke, K.; Heyde, K. Optimal Monte Carlo updating. \textit{Phys. Rev. E} \textbf{2004}, \textit{70}, 056705. 

\bibitem{LOU} Frigessi, A.; Hwang, C. R.; Younes, L. Optimal Spectral Structure of Reversible Stochastic Matrices, Monte Carlo Methods and the Simulation of Markov Random Fields.  \textit{Ann. Appl. Prob.} \textbf{1992}, \textit{2}, 610-628. 

\bibitem{BC sufficiency1} Tierney, L. Markov Chains for Exploring Posterior Distributions.  \textit{Ann. Statist.} \textbf{1994}, \textit{22}, 1701-1728.

\bibitem{BC sufficiency2} Meyn, S. P.; Tweedie, R. L. \textit{Markov Chains and Stochastic Stability}; Springer-Verlag: London, 1993. 

\bibitem{BC sufficiency3} Manousiouthakis, V. I.; Deem, M. W. Strict detailed balance is unnecessary in Monte Carlo simulation. \textit{J. Chem. Phys.} \textbf{1999}, \textit{110}, 2753.

\bibitem{Diaconis} Diaconis, P.; Holmes, S.; Neal, R. M. Analysis of a nonreversible Markov chain sampler. \textit{Ann. Appl. Probab.} \textbf{2000}, \textit{10}, 726-752. 

\bibitem{Chen} Chen, F.; Lov\'asz, L.; Pak, I. Lifting Markov chains to speed up mixing. In \textit{Proceedings of the 31st Annual ACM Symposium on Theory of Computing}, Atlanta, GA, USA, May 1-4, 1999; Association for Computing Machinery: NY, USA, 1999; 275-281. 

\bibitem{Barkema} Schram, R. D.; Barkema,  G. T. Monte Carlo methods beyond detailed balance. \textit{Physica A} \textbf{2015}, \textit{418}, 88-93. 

\bibitem{Ren} Ren, R.; Orkoulas, G. Acceleration of Markov chain Monte Carlo simulations through sequential updating. \textit{J. Chem. Phys.} \textbf{2006}, \textit{124}, 064109. 

\bibitem{Suwa-Todo} Suwa, H.; Todo, S. Markov Chain Monte Carlo Method without Detailed Balance. \textit{Phys. Rev. Lett.} \textbf{2010}, \textit{105}, 120603. 

\bibitem{Suwa-Todo2} Suwa, H.; Todo, S. Geometric allocation approaches in Markov chain Monte Carlo. \textit{J. Phys. Conf. Ser.} \textbf{2013}, \textit{473}, 012013. 

\bibitem{Turitsyn} Turitsyn,  K. S.; Chertkov, M.; Vucelja, M. Irreversible Monte Carlo algorithms for efficient sampling.  \textit{Physica D} \textbf{2011}, \textit{240}, 410-414. 

\bibitem{Weigel} Fernandes,  H. C.; Weigel, M. Non-reversible Monte Carlo simulations of spin models. \textit{Comput. Phys. Commun.} \textbf{2011}, \textit{182}, 1856-1859.

\bibitem{Sakai Hukushima 1D} Sakai, Y.; Hukushima, K. Dynamics of One-Dimensional Ising Model without Detailed Balance Condition. \textit{J. Phys. Soc. Jpn.} \textbf{2013}, \textit{82}, 064003. 

\bibitem{Sakai Hukushima 2D} Sakai, Y.; Hukushima, K. An irreversible Markov-chain Monte Carlo method
with skew detailed balance conditions. \textit{J. Phys. Conf. Ser.} \textbf{2013}, \textit{473}, 012012. 

\bibitem{Sakai Hukushima eigenvalue}  Sakai, Y.; Hukushima, K. Eigenvalue analysis of an irreversible random walk with skew detailed balance conditions. \textit{Phys. Rev. E} \textbf{2016}, \textit{93}, 043318.

\bibitem{Sakai Hukushima simulated tempering} Sakai, Y.; Hukushima, K. Irreversible Simulated Tempering. \textit{J. Phys. Soc. Jpn.} \textbf{2016}, \textit{85}, 104002.

\bibitem{ECMC continuous spins} Michel, M.; Mayer, J.; Krauth,  W. Event-chain Monte Carlo for classical continuous spin models. \textit{EPL} \textbf{2015}, \textit{112}, 20003.

\bibitem{ECMC hard spheres} Bernard, E. P.; Krauth, W.; Wilson, D. B. Event-chain Monte Carlo algorithms for hard-sphere systems. \textit{Phys. Rev. E} \textbf{2009}, \textit{80}, 056704. 

\bibitem{ECMC generalized} Michel,  M.; Kapfer,  S. C.; Krauth,  W. Generalized event-chain Monte Carlo: Constructing rejection-free global-balance algorithms from infinitesimal steps. \textit{J. Chem. Phys.} \textbf{2014}, \textit{140}, 054116.

\bibitem{ECMC heisenberg} Nishikawa,  Y.; Michel, M.; Krauth,  W.; Hukushima, K. Event-chain algorithm for the Heisenberg model: Evidence for $z \simeq 1$ dynamic scaling. \textit{Phys. Rev. E} \textbf{2015}, \textit{92}, 063306.

\bibitem{non reversible parallel tempering} Syed, S.; Bouchard-C\^{o}t\'{e}, A.; Deligiannidis, G.; Doucet, A. 2019, arXiv:1905.02939v2 [stat.CO]. arXive e-prints. https://arxiv.org/abs/1905.02939v2 (accessed: Jan 27, 2020).

\bibitem{hot topic 1} Ichiki,  A.; Ohzeki,  M. Full-order fluctuation-dissipation relation for a class of nonequilibrium steady states. \textit{Phys. Rev. E} \textbf{2015}, \textit{91}, 062105. 

\bibitem{hot topic 2} Ohzeki, M. Stochastic gradient method with accelerated stochastic dynamics. \textit{J. Phys. Conf. Ser.} \textbf{2016}, \textit{699}, 012019. 

\bibitem{hot topic 3} Ichiki, A.; Ohzeki, M. Violation of detailed balance accelerates relaxation. \textit{Phys. Rev. E} \textbf{2013}, \textit{88}, 020101. 

\bibitem{hot topic 4} Ichiki, A.; Ohzeki, M. Langevin dynamics neglecting detailed balance condition. \textit{Phys. Rev. E} \textbf{2015}, \textit{92}, 012105.

\bibitem{hot topic 5} Takahashi,  K.; Ohzeki,  M. Conflict between fastest relaxation of a Markov process and detailed balance condition. \textit{Phys. Rev. E} \textbf{2016}, \textit{93}, 012129. 

\bibitem{hot topic 6} Ohzeki, M.; Ichiki, A. Mathematical understanding of detailed balance condition violation and its application to Langevin dynamics. \textit{J. Phys. Conf. Ser.} \textbf{2015}, \textit{638}, 012003.

\bibitem{hot topic 7} Kaiser, M.; Jack, R. L.; Zimmer,  J. Acceleration of Convergence to Equilibrium in Markov Chains by Breaking Detailed Balance. \textit{J. Stat.Phys.} \textbf{2017}, \textit{168}, 259-287.

\bibitem{REM ST1} Itoh,  S. G.; Okumura, H. Replica-Permutation Method with the Suwa–Todo Algorithm beyond the Replica-Exchange Method. \textit{J. Chem. Theory Comput.} \textbf{2013}, \textit{9}, 570-581.

\bibitem{REM ST2}  Itoh,  S. G.; Okumura, H. Hamiltonian replica-permutation method and its applications to an alanine dipeptide and amyloid‐$\beta$(29–42) peptides. \textit{J. Comput. Chem.} \textbf{2013}, \textit{34}, 2493-2497.

\bibitem{ST Suwa-Todo} Mori, Y.; Okumura, H. Simulated tempering based on global balance or detailed balance conditions: Suwa–Todo, heat bath, and Metropolis algorithms. \textit{J. Comput. Chem.}  \textbf{2015}, \textit{36}, 2344-2349.

\bibitem{krauth book} Krauth, W. \textit{Statistical Mechanics: Algorithms and Computations}; Oxford University Press: New York, 2006.

\bibitem{Hastings} Hastings, W. Monte Carlo sampling methods using Markov chains and their applications. \textit{Biometrika} \textbf{1970}, \textit{57}, 97-109.

\bibitem{Fahim and George} Faizi,  F.; Deligiannidis,  G.;  Rosta, E. in preparation. 

\bibitem{Fahim and George 2}  Faizi,  F.; Deligiannidis,  G.;  Rosta, E. in preparation.  

\bibitem{phase transition 2}  Van Hove, L. Sur L'int\'egrale de Configuration Pour Les Syst\`emes De Particules \'A Une Dimension. \textit{Physica} \textbf{1950}, \textit{16}, 137-143. 

\bibitem{phase transition 3} Ruelle, D. Statistical mechanics of a one-dimensional lattice gas. \textit{Commun. Math. Phys.} \textbf{1968}, \textit{9}, 267-278.

\bibitem{phase transition 4} Ruelle, D. \textit{Statistical mechanics: Rigorous results}; Imperial College Press and World Scientific: London, 1999.

\bibitem{phase transition 5} Cuesta J. A.; S\'anchez, A. General Non-Existence Theorem for Phase Transitions in One-Dimensional Systems with Short Range Interactions, and Physical Examples of Such Transitions. \textit{J. Stat. Phys.} \textbf{2004}, \textit{115}, 869-893. 

\bibitem{ECMC Hard disk MC} Kapfer S. C.; Krauth W. Sampling from a polytope and hard-disk Monte Carlo. \textit{J. Phys. Conf. Ser.} \textbf{2013}, \textit{454}, 012031. 

\bibitem{Glauber Dynamics} Glauber, R. J. Time-Dependent Statistics of the Ising Model. \textit{J. Math. Phys.} \textbf{1963}, \textit{4}, 294.

\bibitem{GCA1} Dress, C.; Krauth, W. Cluster algorithm for hard spheres and related systems. \textit{J. Phys. A} \textbf{1995}, \textit{28}, L597.

\bibitem{GCA2} Liu, J.; Luijten, E. Generalized geometric cluster algorithm for fluid simulation. \textit{Phys. Rev. E} \textbf{2005}, \textit{71}, 066701.

\bibitem{spin glass fail} Coddington, P. D.; Han, L. Generalized cluster algorithms for frustrated spin models. \textit{Phys. Rev. B} \textbf{1994}, \textit{50}, 3058.


\end{thebibliography}
\end{document}